\documentclass[sigplan,screen,10pt]{acmart}

\acmConference[PL'19]{ACM SIGPLAN Conference on Programming Languages}{January 01--03, 2019}{New York, NY, USA}
\acmYear{2019}
\acmISBN{} 
\acmDOI{} 
\startPage{1}

\usepackage{booktabs}   
\usepackage{subcaption} 

\usepackage{listings, listings-rust}
\usepackage{amsfonts}
\usepackage{stmaryrd}
\usepackage{mathtools}
\usepackage{amsmath}
\usepackage{tikz}
\usetikzlibrary{arrows}
\usetikzlibrary{calc}
\usepackage{calc} 
\usepackage{epigraph}
\usepackage[frozencache,cachedir=.,outputdir=.]{minted}
\usepackage{newfloat}
\usepackage{syntax}
\usepackage{graphicx}
\usepackage{fancyvrb}
\usepackage{pdfpages}
\usepackage{pict2e}
\usepackage{verbatim}
\usepackage{ragged2e}
\usepackage{adjustbox}
\usepackage{hyperref}
\usepackage{balance}

\usepackage{amsmath,amssymb,trimclip,adjustbox}

\newcommand\fnurl[2]{%
\href{#2}{#1}\footnote{\url{#2}}%
}

\tikzstyle{every picture}+=[remember picture]

\DeclareMathOperator{\Pound}{\#}
\DeclareMathOperator{\Align}{@~}
\DeclareMathOperator{\HasType}{\textrm{::}~}
\DeclareMathOperator{\OR}{{\bf \parallel}}
\newcommand{\Exists}[2]{\exists~#1~.~#2}
\newcommand{\Con}[2]{\textrm{Con } #1~#2}

\newcommand{\lbparen}{%
  \mathopen{~%
    \sbox0{$()$}%
    \setlength{\unitlength}{\dimexpr\ht0+\dp0}%
    \raisebox{-\dp0}{%
      \begin{picture}(.32,1)
      \linethickness{\fontdimen8\textfont3}
      \roundcap
      \put(0,0){\raisebox{\depth}{$($}}
      \polyline(0.32,0)(0,0)(0,1)(0.32,1)
      \end{picture}%
    }%
  ~}%
}

\newcommand{\rbparen}{%
  \mathclose{~%
    \sbox0{$()$}%
    \setlength{\unitlength}{\dimexpr\ht0+\dp0}%
    \raisebox{-\dp0}{%
      \begin{picture}(.32,1)
      \linethickness{\fontdimen8\textfont3}
      \roundcap
      \put(-0.08,0){\raisebox{\depth}{$)$}}
      \polyline(0,0)(0.32,0)(0.32,1)(0,1)
      \end{picture}%
    }%
  ~}%
}

\def\<#1>{\synt{#1}}

\DeclareFloatingEnvironment[
  fileext  = logr,
  listname = {List of Grammars},
  name     = Grammar,
  placement = htp
]{Grammar}

\def \ourlang{Floorplan}
\def \ourlangL{floorplan}
\def \ourlangExt{flp}

\newcounter{codeC}
\newcommand{\rcodeC}[1]{F\refstepcounter{codeC}\label{#1}}
\newcommand{\newCodeC}[1]{(\rcodeC{#1}\arabic{codeC})}

\newcounter{codeR}

\newcounter{codeCore}

\newcounter{eqn}[section]
\newcommand{\reqn}[1]{\refstepcounter{eqn}\label{#1}}
\newcommand{\newEqn}[1]{(\reqn{#1}\arabic{section}.\arabic{eqn})}
\newcommand{\refEqn}[1]{\arabic{section}.\ref{#1}}

\newcounter{deftn}[section]

\newcounter{lmma}[section]

\newcounter{thm}[section]

\def \ourlangLines {47~}
\def \pointerArithmeticLines {67~}
\def \offsetConstantLines {25~}
\def \bitmapLines {169~}
\def \generatedLines {877~}
\newmintinline[ourmint]{\ourlangL}{}
\newmintinline[rustInline]{rust}{}
\def \mintFontSize {\small}
\newminted{\ourlangL}{gobble=2,fontsize=\mintFontSize}

\newminted{rust}{gobble=2,fontsize=\mintFontSize}
\newminted{layoutFLP}{xleftmargin=1.7em,linenos=true,fontsize=\mintFontSize,breaklines=true}

\begin{document}

\title[Spatial Layout in MM Systems]{Floorplan: Spatial Layout \protect\\ in Memory Management Systems}         


\author{Karl Cronburg}
\orcid{0000-0002-8338-8789}
\affiliation{
  \institution{Tufts University}            
}
\email{karl@cs.tufts.edu}          

\author{Samuel Z. Guyer}
\affiliation{
  \institution{Tufts University}
}
\email{sguyer@cs.tufts.edu}


\begin{abstract}
In modern runtime systems, memory layout calculations are hand-coded in systems
languages. Primitives in these languages are not powerful enough to describe
a rich set of layouts, leading to reliance on ad-hoc macros, numerous
interrelated static constants, and other boilerplate code. Memory management
policies must also carefully orchestrate their application of address
calculations in order to modify memory cooperatively, a task ill-suited to
low-level systems languages at hand which lack proper safety mechanisms.

In this paper we introduce Floorplan, a declarative language for specifying
high level memory layouts. Constraints formerly implemented by describing
how to compute locations are, in Floorplan, defined declaratively using
explicit layout constructs. The challenge here was to discover constructs
capable of sufficiently enabling the automatic generation of address
calculations. Floorplan is implemented as a compiler for generating a Rust
library. In a case study of an existing implementation of the immix garbage
collection algorithm, Floorplan eliminates 55 out of the 63 unsafe lines of
code: 100\% of unsafe lines pertaining to memory safety.

\end{abstract}

%



\maketitle


\newpage
\section{Introduction} \label{sec:introduction}

The design of a memory manager is often hidden away in the runtime system and
rarely discussed the way more prominent language features, such as syntax and
semantics, are. A number of factors contribute to this state of affairs. First,
each implementation of a managed language typically has its own memory
manager, built from scratch, resulting in an almost total absence of shared
code. Second, runtime system code is difficult to comprehensively
understand: low-level and intricate, with a premium placed on performance.
Finally, crucial design elements are often buried in the code, such as in
simple yet pervasive pointer arithmetic and bitwise manipulations. These
operations have ramifications on design
elements across the entire system. As a result, these design elements are
intrinsically hard to get correct the first time, and hard to diagnose when
they are incorrect. Without a specification of these design elements,
properties of a memory management algorithm are difficult or impossible to
check and reason about formally. Documentation, when present, is in the form of
informal and often inaccurate or ambiguous comments. Traditional memory safety
tools~\cite{Nethercote:2007:Valgrind}~fall short because they typically assume
that the memory allocator is allocating memory correctly in the first place.

In this work we take a first step toward remedying this situation: we
present a declarative, domain-specific language (DSL), called \emph{\ourlang},
for describing the structure of a heap as laid out by a memory manager.
\ourlang~is inspired by PADS~\cite{pads}, a language for describing ad hoc data
file formats. A \ourlang~specification looks like a grammar, augmented with
memory management specific features.
\ourlang~provides powerful ways to specify the sizes, alignments, and
relationships among chunks of memory, resulting in very compact descriptions.
The key idea is that any correct state of the heap can be represented as a
string (a sequence of bytes or tokens) derivable from a \ourlang~grammar.
Grammars are a natural choice because they match the configuration of most
modern memory managers, which comprise layers of code that carve up memory into
smaller and smaller pieces.
Every~\cite{scalloc},
\cite{JikesRVM},
\cite{Hoard2000},
\cite{HeapLayers},
\cite{immix},
\cite{supermalloc},
\cite{dlmalloc},
\cite{rust-gc},
\cite{GHC-GC},
\cite{openjdk-hotspot},
memory manager we've studied exhibits this
allocation scheme.

Note that \ourlang~does not attempt to capture the policy details of any
particular memory management algorithm. The closest \ourlang~gets to capturing
policy details is in its ability to logically connect multiple pieces of memory,
e.g. a bit map representing allocated cells in a block.
The \ourlang~compiler generates the low-level mechanisms
\textemdash~pointer calculations, bit masking, etc. \textemdash~that the developer
calls in order to implement some memory management policy.
For example, the \ourlang~compiler
automates the synthesis of constants and pointer calculations for accessing an
object liveness bitmap while saying nothing about how liveness or reachability are computed.
Ongoing future work aims to leverage \ourlang~specifications to debug
algorithmic errors resulting in memory corruption.
Such temporal errors are not easily detectable with frameworks like
Valgrind~\cite{Nethercote:2007:Valgrind} and PIN~\cite{Luk:2005:Pin}.  These
frameworks can be leveraged more methodically with a layout specification
language like \ourlang.


\subsection{Contributions}

To summarize, this work makes the following contributions:

\begin{itemize}
  \item A declarative specification language based in part on a novel formalization of
    union types in Section~\ref{sec:semantics}. Floorplan allows users to express a
    memory layout as a \emph{specification}, defining the spatial
    relationships among one or more system-defined types of memory.
  \item The
    \ourlang~specification of the layout of a state-of-the-art garbage
    collection algorithm: immix as implemented in Rust.
  \item Formal rules for translating surface syntax to a core expression language,
    and a denotational semantics for how toreduce a  memory layout to a set of trees with
    bytes at their leaves.
  \item A \ourlang~compiler targeting Rust.
  \item Boilerplate reduction and memory safety results from integrating a
    \ourlang~specification with the Rust implementation of immix~\cite{rust-gc}.
\end{itemize}

\section{Motivation} \label{sec:motivation}

\begin{figure}
\lstset{basicstyle=\ttfamily\scriptsize}
\begin{lstlisting}[language=Java]
int SCALAR_HEADER_SIZE =
  JAVA_HEADER_BYTES + OTHER_HEADER_BYTES;
int ARRAY_HEADER_SIZE =
  SCALAR_HEADER_SIZE + ARRAY_LENGTH_BYTES;
/** offset of object reference
    from the lowest memory word */
Offset TIB_OFFSET = JAVA_HEADER_OFFSET;
Offset STATUS_OFFSET = TIB_OFFSET.plus(STATUS_BYTES);
Offset AVAILABLE_BITS_OFFSET =
  VM.LittleEndian ?
      STATUS_OFFSET
    : STATUS_OFFSET.plus(STATUS_BYTES - 1);
int HASH_CODE_SHIFT = 2;
Word HASH_CODE_MASK =
  Word.one()
    .lsh(10)
    .minus(Word.one())
    .lsh(HASH_CODE_SHIFT);
/** How many bits are allocated to a thin lock? */
int NUM_THIN_LOCK_BITS = ADDRESS_BASED_HASHING ? 22 : 20;
/** How many bits to shift to get the thin lock? */
int THIN_LOCK_SHIFT = ADDRESS_BASED_HASHING ? 10 : 12;
\end{lstlisting}
\rule{\linewidth}{.5pt}
\caption{\label{fig:jikes-rvm-constants} Code fragment from Jikes RVM~\cite{JikesRVM} showing
some of the Java header related constants.}
\end{figure}

Spatial layout is fundamental to the problem of dynamic memory
management. Memory managers employ a variety of layout schemes to carve
up raw memory, and each scheme is influenced by the particular
algorithm being implemented.
Great care goes into designing a layout which permits highly efficient operation
of crucial layout operations.
For example, a generational garbage collector might be
laid out such that the nursery is in a lower part of memory than the
older space. This choice allows the write barrier to be implemented
exclusively with address comparisons. Similarly, a free-list allocator
might divide pages into cells of equal size, like an array, with a bit
map of free cells at the start of each page. This design allows the
meta-data to be found by simply masking off the low bits of any cell
address; the corresponding bit can then be computed easily by dividing
the low bits of the address by the cell size. These
optimizations improve performance, but are
only valid if the layout permits them.

\smallskip\noindent
{\bf Software maintenance.}
In all the memory managers we've studied, spatial layout is only formally
expressed by the code that implements it.
Figure~\ref{fig:jikes-rvm-constants} is a typical example, taken from
MMTk~\cite{mmtk}, the memory management toolkit.
Notice, in particular, the calculation of the hash code mask -- clearly, great
care is required to write, modify, and maintain such code.
While MMTk is among the most
meticulously engineered of memory mangers, this memory manager
consists of boilerplate code in excess of $2,399$ lines of address arithmetic
calulations as per the following bash command:

\begin{verbatim}
$ find MMTk/ rvm/ -name *.java -exec egrep \
-e "\.(one|lsh|plus|minus|rshl|and|EQ)\(" \
-e "\.(zero|isZero|diff|store|load)\(" "{}" \; \
| wc -l
2399
\end{verbatim}

\smallskip\noindent
{\bf Static typing.}
A common problem in the memory management field occurs when a memory manager
is implemented with generic pointer types exhibiting memory-related bugs.
Often such generic pointers are distinguishable based on their value.
Suspicious pointer values are manually detected based on intrinsic properties,
such as alignment checks and assertions pertaining to relationships with
other known in-memory structures, e.g. containing blocks and regions.

Complicating matters further, specialized pointer types need not even be
distinguishable from one another by value dynamically. For example
the Rust code in Figure~\ref{fig:rust-motivation} shows code implementing a
block of memory cells.
This code operates over memory with
raw address types, and consists of numerous address calculations on generic
pointer types.
By this design, the address of the start of a block is
the same as the address of the first cell in that
block. Code that uses this class could call either method -- it does not matter
which. If the layout changed, though, for example by adding a bit map, then the
methods \emph{would} be different. Code that calls the \verb|block.start()|
method expecting a pointer to a cell would now fail at runtime in confusing
ways. These failures motivate the approach of generating specialized address
types.


\begin{figure}
\lstset{basicstyle=\ttfamily\small}
\begin{lstlisting}[language=Rust]
#[repr(C)]
#[derive(Copy, Clone, Eq, Hash)]
pub struct Address(usize);
pub struct Block { start : Address }
impl Block {
  pub fn start(&self) -> Address {
    self.start }
  pub fn first_cell(&self) -> Address {
    self.start }}
\end{lstlisting}
\caption{\label{fig:rust-motivation} Abbreviated snippet of Rust code
from an implementation of immix~\cite{rust-gc}.}
\end{figure}

\smallskip\noindent
{\bf Dynamic typing.}
In a runtime system the configuration of the heap changes over time in highly
mechanistic, and largely superficial, ways. For example, the heap's configuration
changes when a piece of addressable memory in the heap changes type. This
results in
new offset and address calculations being allowed on that address. These
calculations are used to implement various allocation schemes which
\emph{combine}, \emph{carve up}, or \emph{interchange} pieces of memory.

Some allocation schemes \emph{combine} multiple operating system level pieces
of memory into larger pieces of memory. For example, multiple contiguous pages can be
combined to form a single block. This combination is typically implemented with a
simple multiplication or bit-shifting operation.

Other allocation schemes \emph{carve} a single piece of memory into
multiple subcomponents. For example, a block may be carved up into
cells, with a bitmap at the beginning of the block. Carving up of memory
is typically implemented with a simple offset added to an address, and a
subsequent bounds check address comparison to detect block overflow.

Finally some allocation schemes define two or more pointer types to be
\emph{interchangeable}. For example, a cell of memory is either allocated or free,
with differing internal layouts. A free cell controlled by a doubly-linked
list policy typically contains two pointers. Accesses to a free cell must
therefore be implemented with an offset addition to the cell's base address.
Such core layout operations are simple in isolation, yet the design choices
describing their composition are complex.

\smallskip\noindent
{\bf Efficiency.}
Address calculations need to behave such that an amortized analysis
of an allocation scheme yields a highly efficient implementation.
Existing handwritten calculations exhibit this efficiency, so
generating address calculations to semantically and stylistically match
handwritten code makes sense.
Precise control over the form of generated code ensures efficiency-motivated
size, alignment, and padding invariants hold. Generating code
also forgoes the manual writing of numerous lines of stylistically
similar code.
Existing memory managers lack precise and formal specifications
of their memory layouts. Memory managers can benefit from support for
various forms of analysis, debugging, and code generation which this work tackles.
In this paper we take a generative approach: we describe a specification
language, its translation to a core calculus, and a compiler for generating
Rust code.

\section{Language overview with examples} \label{sec:syntax}


\begin{figure}
  \begin{tabular}{rll}
  {\bf Code}        & {\bf Nonterminal}     & {\bf Explanation} \\ \hline
    \ourmint{||sz||}      & \synt{mag}         & ``has size \ourmint{sz}'' \\
    \ourmint{@(sz)}       & \synt{align}         & ``\ourmint{sz} address alignment'' \\
    \ourmint{@|sz|@}      & \synt{magAlign}         & ``same \synt{align} and \synt{mag}'' \\
    \ourmint{# Bar}       & \synt{demarc-val}      & ``some number of \ourmint{Bar}s'' \\
    \ourmint{foo : Bar}   & \synt{field}        & ``field \ourmint{foo} contains a \ourmint{Bar}'' \\
    \ourmint{Bar, Baz}    & \synt{seq}             & ``\ourmint{Bar} followed by \ourmint{Baz}'' \\
    \ourmint{Foo -> Bar}  & \synt{layer}              & ``\ourmint{Foo} consists of \ourmint{Bar}'' \\
    \ourmint{Bar | Baz}   & \synt{union}             & ``one of \ourmint{Bar} or \ourmint{Baz}'' \\
    \ourmint{FOO | BAR}   & \synt{enum}       & ``in state \ourmint{FOO} or \ourmint{BAR}'' \\
  \end{tabular}
  \caption{\label{fig:semantics} Informal semantics of
  constructs and operators in \ourlang. \ourmint{Bar} and \ourmint{Baz}
  represent arbitrary \synt{demarc-val} values, \ourmint{FOO} and \ourmint{BAR}
  represent state flags of an \synt{enum}, \ourmint{Foo} represents an
  identifier, and \ourmint{sz} is some \synt{size-arith}.
  } 
\end{figure}


The most fundamental operation in memory management is to take an unstructured
piece of memory and to give it structure through \emph{demarcation}.
Demarcation is the dividing up of a layer of memory into a partitioning of components.
Multiple layers of memory form an allocation hierarchy.

In order to allocate a piece of memory, a memory manager tracks metadata distinguishing
a free piece from the same allocated piece. The state of this piece of memory, free or
allocated, determines its layout. Existing systems written in C model this behavior with
unions. For example, the first word of a free-list based allocator's free piece
might contain a pointer, while that same word of memory once allocated might
contain an object header. In order to access this allocated object's payload, a
memory manager calculates the payload's offset from the base of the containing
piece of memory. The ordering of fields in this piece of memory, header and
payload, define its layout. Existing systems often model the ordering of fields
with offset constants. For example, a memory manager computes the location of
a payload in terms of the size of its header. In this section, we introduce
\ourlang~with similarly motivated examples.

Grammar $1$, below, through Grammar $4$ specify the syntactic
constructs of a \ourlang~specification in EBNF form. For a quick reference
guide on how to read a \ourlang~specification, refer to
Figure~\ref{fig:semantics}. The grammars below are inline figures, which we
recommend inspecting in the order they are presented before reading the
remainder of this section.

\noindent%
\rule{\linewidth}{.5pt}
\noindent%
\justify
{\bf Grammar 1}: Literal lexemes. Layers \& fields are types,
formals represent natural numbers, and flags for enums.\\

\noindent%
\begin{minipage}[t]{0.55\columnwidth}
\begin{grammar}
<layer-id>   ::= [A-Z][a-zA-Z\_]*

<field-id>  ::= [a-z][a-zA-Z\_]*

<formal-id> ::= [a-z][a-zA-Z\_]*

<flag-id>   ::= [A-Z][A-Z\_]*
\end{grammar}
\end{minipage}%
\begin{minipage}[t]{0.45\columnwidth}
\begin{grammar}
<literal> ::= <bin> | <int>

<bin>       ::= 0b[01]+

<int>       ::= [0-9]+

<prim> ::= \lit{bits} | \lit{bytes} | \lit{words} | \lit{pages}
\end{grammar}
\end{minipage} \\
%

\grammarindent2.5cm
\noindent%
\rule{\linewidth}{.5pt}
\noindent%
\justify
{\bf Grammar 2:} Arithmetic language for memory sizes.


\begin{grammar}
<lit-arith> ::=
       <literal> | \lit{(} <lit-arith> `)'
  \alt <lit-arith> <lit-arith-op> <lit-arith>

<lit-arith-op> ::= \lit{$+$} | \lit{$-$} | \lit{$*$} | \lit{$/$}
  | \lit{\^{}}

<size-arith> ::=
       <lit-arith>? <prim> | \lit{(} <size-arith> \lit{)}
  \alt <size-arith> <size-arith-op> <size-arith>

<size-arith-op> ::=
     \lit{$+$}
  |  \lit{$-$}

\end{grammar}

%

\noindent%
\rule{\linewidth}{.5pt}
\noindent%
\justify
  {\bf Grammar 3:} Layers of memory with annotated magnitudes, alignments,
  simultaneous annotations (\synt{magAlign}), scoped formal parameter declarations,
  and containment (\synt{contains}) compiler annotation hints\footnote{These
  instruct the compiler to generate functions for converting to the containing
  \synt{layer-id} and vice-versa when memory alignments permit.}.

\begin{grammar}
<layer-simple> ::= <layer-id> (\lit{\textless} <formals> \lit{\textgreater})?
  (<mag>? <align>? | <magAlign>?) <contains>* \lit{-\textgreater} <demarc-val>

<layer> ::=      <layer-simple> | \lit{(}  <layer-simple> \lit{)}

<mag> ::= \lit{||} <size-arith> \lit{||}

<align> ::= \lit{@} \lit{(} <size-arith> \lit{)}

<magAlign> ::= \lit{@|} <size-arith> \lit{|@}

<formals> ::= <formal-id> (\lit{,} <formal-id>)* `,'?

<contains> ::= \lit{contains} `(' <layer-id> `)'
\end{grammar}



\noindent%
\rule{\linewidth}{.5pt}
\noindent%
\justify
{\bf Grammar 4:} Demarcatable atomic units of memory.
\grammarindent2cm
\begin{grammar}
<demarc-val> ::= (`#' | <formal-id>)? (<enum> | <bits> | <union>
\alt <seq> | <ptr> | <size-arith> | <macro>)

<seq> ::= \lit{seq} \lit{\{} <demarc> (`,' <demarc>)* `,'? \lit{\}}

<union> ::= \lit{union} \lit{\{} <demarc> (`|' <demarc>)* `|'? \lit{\}}

<demarc> ::= <field> | <layer> | <demarc-val>

<field> ::= <field-id> \lit{:} <demarc-val>

<ptr> ::= (<layer-id> | <field-id>) `ptr'

<enum> ::= \lit{enum} \lit{\{} <flag-id> (`|' <flag-id>)* `|'? \lit{\}}

<bits> ::= \lit{bits} \lit{\{} <bits-exp> (`,' <bits-exp>)* `,'? \lit{\}}

<bits-exp> ::= <field-id> \lit{:} <size-arith>

<macro> ::= <layer-id> (`<' <args> `>')?

<arg> ::= <formal-id> | <literal>

<args> ::= <arg> (`,' <arg>)* `,'?
\end{grammar}
\noindent%
\rule{\linewidth}{.5pt}

\subsection{What is a Floorplan demarcation} \label{sec:flp-demarcation}

In Grammar~4 we introduced the syntactic form for the notion of a demarcation.
A \emph{demarcation} is a partitioning\footnote{Including finitely many
partitions of size zero.} of a layer of a heap. A boundary position in memory
defining the partition of two or more \synt{layer} and \synt{field} types may
(and often does) coincide with another layer's boundary.

For instance in our block-containing-cells motivating example (Figure~\ref{fig:rust-motivation}) the beginning
boundary of a block coincides with the boundary of that block's
first cell. We can encode this memory layout as follows:

\begin{tabular}{p{0.80\columnwidth} p{0.20\columnwidth}}
\begin{floorplancode}
  Cell -> seq { Header  -> 1 words,
                Payload -> 7 words }
  Block ||2^16 bytes|| -> # Cell
\end{floorplancode}
& ~\linebreak\linebreak \newCodeC{code:cell-and-block}
\end{tabular}

This code declares a block of cells with total size
$2^{16}$ bytes. The ``\#'' operator indicates that the \ourmint{Cell} declaration
should be repeated as many times as necessary in order to exactly fill the total size.
The \ourmint{Cell} reference on the last line of F\ref{code:cell-and-block} parses
as a \synt{macro} expression\footnote{Macros are not formally specified: they
are a pre-processing pass to the compiler. Recursive macros are forbidden.} which must reference a top-level \synt{layer-id}
declaration of the specification file (\texttt{.flp} filename extension). A \synt{macro}
expression is syntactically replaced with its corresponding declaration.


From the layout in F\ref{code:cell-and-block} the compiler generates specialized address
types for pointers to a \ourmint{Cell}, \ourmint{Header}, \ourmint{Payload},
and \ourmint{Block} respectively. For safety reasons, a memory manager must
only be able to cast from a \ourmint{Block} address to a \ourmint{Cell}
address and not to, say, a \ourmint{Payload} address. Therefore the compiler
generates (simplified here) Rust code identical in purpose to that of
Figure~\ref{fig:rust-motivation}:

\begin{tabular}{p{0.80\columnwidth} p{0.20\columnwidth}}
\begin{center}
  {\bf Types \& casts generated for Code~F\ref{code:cell-and-block}}
\end{center}
\lstset{language=Rust,basicstyle=\footnotesize\ttfamily}
\begin{lstlisting}
pub struct CellAddr(usize);
pub struct HeaderAddr(usize);
pub struct PayloadAddr(usize);
pub struct BlockAddr(usize);
impl BlockAddr {
  pub fn get_first_cell(&self) -> CellAddr {
    CellAddr::from_usize(self.as_usize()) } }
\end{lstlisting}
  & ~\linebreak\linebreak\linebreak\linebreak\linebreak (R\refstepcounter{codeR}\label{code:rust:cell-and-block}\arabic{codeR})
\end{tabular}

While this code is implementable by hand, the complier systematically enforces
which conversions are memory-safe. Memory-safety in Floorplan is heavily
influenced by where coinciding boundaries occur. These occur wherever two
\synt{layer} or \synt{field} declarations are nested inside of one another
under one condition: the nested path traverses neither the tail of a \synt{seq}
nor \synt{demarc-val} annotated with a repetition\footnote{More on \synt{\#}
and \synt{formal-id} repetitions four paragraphs from here.}. Under this
condition, \ourlang~semantics (Section~\ref{sec:semantics}) guides the compiler
in generating safe address conversions. Statically unsafe conversions
are disallowed by construction.

\subsection{Implementing bit-fields and repetitions}

A header word on an object in a memory manager
typically relies on intricately implemented offset constants to function,
like back in Figure~\ref{fig:jikes-rvm-constants}. For example, we might
want to modify the \ourmint{Header} portion of Code F\ref{code:cell-and-block}
to support bit-level manipulation in a traditional mark-sweep
garbage collector:

\begin{tabular}{p{0.80\columnwidth} p{0.20\columnwidth}}
\begin{floorplancode}
  Header @|1 words|@ -> bits {
    MARK : 1 bits, REF : 7 bits,
    UNUSED  : (1 words - 1 bytes) }
\end{floorplancode}
& ~\linebreak\linebreak \newCodeC{code:cell-header}
\end{tabular}

First, Code F\ref{code:cell-header} constrains the alignment of header
words to start on a \ourmint{@|1 words|@} boundary. In addition, the memory
manager needs to be able to access (read and write) the contents of the
\ourmint{MARK} and \ourmint{REF} bits in order to mark and record the location
of pointers in the payload, respectively. To facilitate this requirement,
the compiler generates, e.g., the following constants and accessors:

\vspace{-2mm}
\begin{tabular}{p{0.80\columnwidth} p{0.20\columnwidth}}
\begin{center}
  {\bf Offset constants generated from Code F\ref{code:cell-header}}
\end{center}
\lstset{language=Rust,basicstyle=\footnotesize\ttfamily}
\begin{lstlisting}
struct HeaderAddr(usize);
impl HeaderAddr {
  pub const MARK_LOW_BIT : usize = 0;
  pub const MARK_NUM_BITS : usize = 1;
  pub const MARK_MASK : u8 = 0b00000001;
  pub const REF_LOW_BIT : usize = 1;
  pub const REF_NUM_BITS : usize = 7;
  pub const REF_MASK : u8 = 0b11111110;
  pub fn set_MARK_bit(&self, val: bool) {
    self.store::<u8>(val as u8) }
  pub fn get_MARK_bit(&self) -> bool) {
    (self.load::<u8>() as bool) } }
\end{lstlisting}
& ~\linebreak\linebreak\linebreak\linebreak\linebreak\linebreak\linebreak (R\refstepcounter{codeR}\label{code:rust:cell-header}\arabic{codeR})
\end{tabular}
\vspace{-4mm}

Furthermore, a memory manager must be able to allocate pointers in the payload
and mark their location in the \ourmint{REF} field.\footnote{How the runtime determines
which \ourmint{REF} bit marks which payload word is outside the scope of this work.}
For example, the layout can dictate that pointer fields in
an application object comprise the first $n$ words of the payload by
replacing the \ourmint{Payload} in F\ref{code:cell-and-block} with:

\vspace{-2mm}
\begin{tabular}{p{0.80\columnwidth} p{0.20\columnwidth}}
\begin{floorplancode}
  Payload ||7 words|| -> seq {
    refs: # (Cell ptr), rem:  # (1 words) }
\end{floorplancode}
& ~\linebreak \newCodeC{code:cell-payload}
\end{tabular}
\vspace{-4mm}

Notice here that the two ``\#'' operators act together to fill the necessary
space (7 words) available to them. Code F\ref{code:cell-payload} denotes $8$
distinct layouts: the number of permutations by which two natural numbers can
sum to $7$. These permutations include $(0$ pointers, $7$ words$)$, $(1$ pointer,
$6$ words$)$, and so on until $(7$ pointers, $0$ words$)$. In order to allocate
some number of pointers, the compiler needs to give us a way to (1) access the
\texttt{refs} field of a \texttt{Payload}, (2) initialize a pointer to the
\texttt{rem} field, and (3) allocate an additional cell pointer. Code
R\ref{code:rust:cell-payload} below exhibits these functions:

\vspace{-2mm}
\begin{tabular}{p{0.82\columnwidth} p{0.18\columnwidth}}
\begin{center}
  {\bf Allocation pattern generated from Code F\ref{code:cell-payload}}
\end{center}
\lstset{language=Rust,basicstyle=\footnotesize\ttfamily}
\begin{lstlisting}
impl PayloadAddr {
  pub fn cast_payload_to_refs(&self)
    -> RefsAddr { // #1
      RefsAddr(self.as_usize()) }
  pub fn init_rem_after_refs(p1: RefsAddr
    , bytes: usize) -> RemAddr { // #2
      debug_assert!(bytes%BYTES_IN_POINTER==0);
      p1.plus::<RemAddr>(bytes) }
  pub fn bump_new_Cell_ptr(rhs: RemAddr)
    -> (CellAddr, RemAddr) { // #3
      (rhs.plus(0),rhs.plus(BYTES_IN_POINTER)}}
\end{lstlisting}
& ~\linebreak\linebreak\linebreak\linebreak\linebreak\linebreak (R\refstepcounter{codeR}\label{code:rust:cell-payload}\arabic{codeR})
\end{tabular}
\vspace{-4mm}

Take for granted that we have access to the \texttt{PayloadAddr} of some cell.
Function \texttt{\#1} above accesses the \texttt{refs} field of our payload. From
this address we can initialize, with \texttt{\#2}, the remainder (\texttt{rem})
to start zero bytes after the start of the payload. With \texttt{\#3} we can
then allocate a new pointer with our \texttt{RemAddr} returned by \texttt{\#2}.
To allocate more pointers we iterate as necessary over \texttt{\#3}, because
\texttt{\#3} returns an updated \texttt{RemAddr}. The compiler knows to
generate this allocation pattern because two adjacent fields each contain a
repetition.

\subsection{Implementing union types}

In contrast to Code~F\ref{code:cell-payload}, we might want a more permissive
object field layout where pointer fields can appear in any order in the
payload. For example:

\vspace{-2mm}
\begin{tabular}{p{0.80\columnwidth} p{0.20\columnwidth}}
\begin{floorplancode}
  Payload ||7 words|| ->
    # union { Cell ptr | (1 words) }
\end{floorplancode}
& ~\linebreak \newCodeC{code:cell-payload2}
\end{tabular}
\vspace{-4mm}

\refstepcounter{codeR}

In F\ref{code:cell-payload2}, the ``\#'' operator acts to fill precisely $7$
words of memory. In doing so, this particular ``\#'' operates equivalently
to the POSIX Extended Regular Expression (ERE) limited repetition
expression \texttt{(a|b)\{7\}}. As with this regex, F\ref{code:cell-payload2}
denotes $2^7 = 128$ distinct layouts: the number of permutations (with repeats)
of the elements of the union fitting into $7$ words. If instead we made a typo
and wrote \ourmint{(10 words)} in place of \ourmint{(1 words)}, the compiler
reports to us a consistency warning: the \ourmint{(10 words)} branch of the
union in F\ref{code:cell-payload2} is dead code which does not contribute
to a valid payload.

\subsection{Implementing lookup tables}

A memory manager often relies on metadata in lookup tables and byte maps.
To indicate the relationship between metadata and memory
it describes, the same \synt{formal-id}, \texttt{cnt}, can logically link two or more
pieces of memory:



\vspace{-2mm}
\begin{tabular}{p{0.80\columnwidth} p{0.20\columnwidth}}
\begin{floorplancode}
  Cell<sz> -> sz (1 words)
  SizeKls<sz, cnt> @|2^16 bytes|@ -> seq {
    cells: cnt Cell<sz>, map: cnt (1 bytes) }
  Kls16 -> SizeKls<16>
\end{floorplancode}
& ~\linebreak\linebreak \newCodeC{code:size-class}
\end{tabular}
\vspace{-4mm}

Code F\ref{code:size-class} implements a 16-word size-class block
of memory, with a byte map at the end of each block. Note that macro expressions are
curried, so only the first argument need be expanded on the last line above. The
compiler generates functions capable of translating between a cell and its corresponding
byte entry in the \ourmint{map}. For example in order to update the byte entry
of some cell, we can call the set function in Code~R\ref{code:rust:size-class} on that
cell's address, along with the value we want the map to remember:

\vspace{-2mm}
\begin{tabular}{p{0.80\columnwidth} p{0.20\columnwidth}}
\begin{center}
  {\bf Mapping code generated from Code F\ref{code:size-class}}
\end{center}
\lstset{language=Rust,basicstyle=\footnotesize\ttfamily}
\begin{lstlisting}
pub struct Cell_16([usize; 16]);
pub struct Cell2Byte {
  pub fBase: CellAddr, // from
  pub tBase: ByteAddr, // to
  pub e: Kls16EndAddr }
impl Cell2Byte {
  pub fn set(&self, fA: CellAddr, val: u8) {
  debug_assert!(fA >= self.fStart);
  let idxV = (fA - self.fBase) >> 7;
  let loc = self.tBase.offset::<Cell_16>(idxV);
  debug_assert!(self.e > loc); loc.store(val); } }
\end{lstlisting}
& ~\linebreak\linebreak\linebreak\linebreak\linebreak (R\refstepcounter{codeR}\label{code:rust:size-class}\arabic{codeR})
\end{tabular}

Dozens more functions are generated alongside \texttt{set()} in Code R\ref{code:rust:size-class}.
We struggled to define meaningful naming schemes for generated address-types.
For example \texttt{Cell_16} above comes from the \synt{macro} expressions on the third
and fourth lines of Code F\ref{code:size-class}.
Code R\ref{code:rust:size-class} also exemplifies generated debugging assertions. Again,
while these assertions can be manually written, formally deriving the largely
trivial ones such as these bounds checks is feasible.

\section{Study: Immix in Rust} \label{sec:immix-in-rust}



\begin{figure}
  

  \includegraphics[width=\columnwidth]{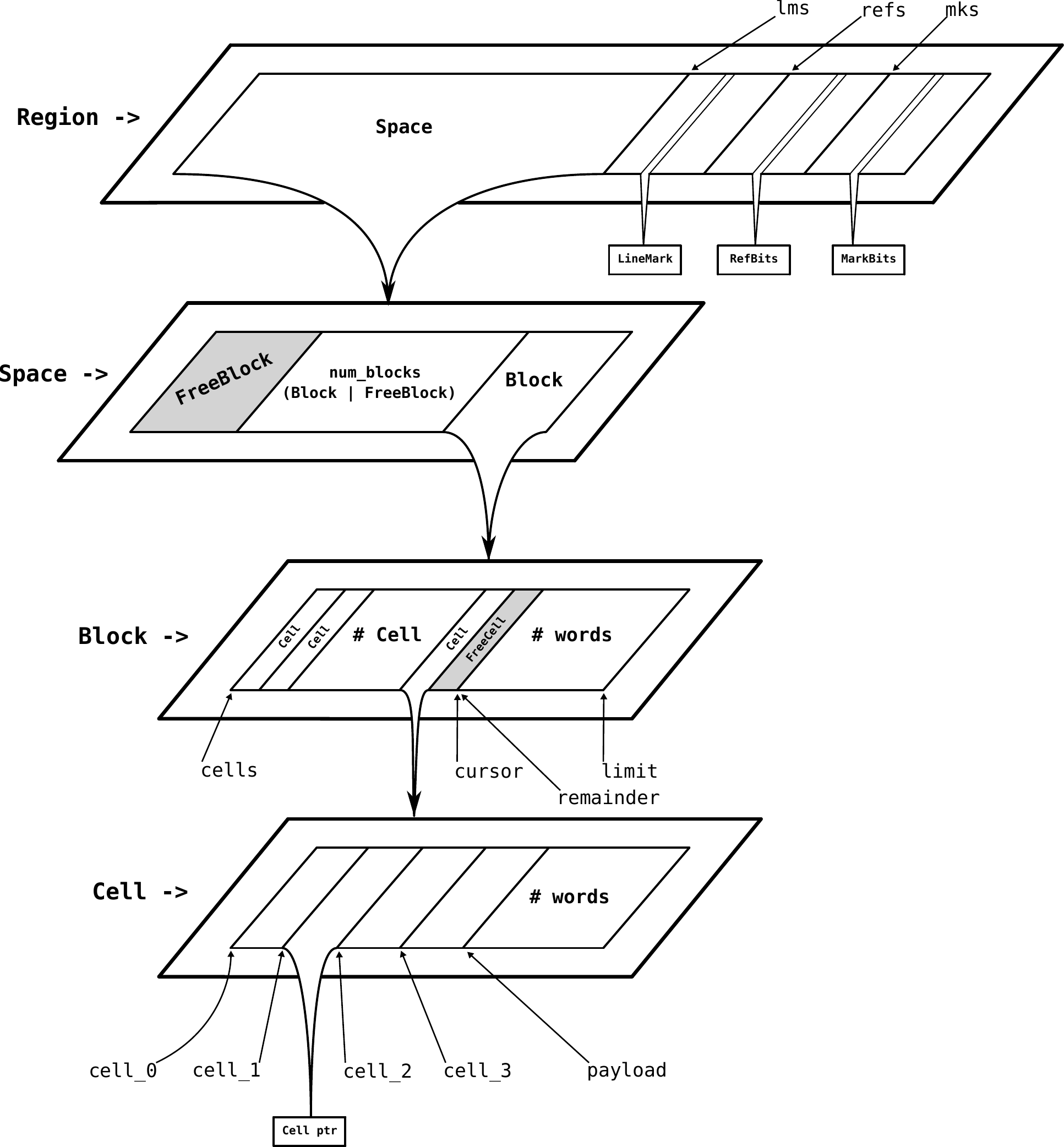}

  \caption{\label{fig:demarc-region} A four-layered demarcation diagram depicting the immix-rust
  layout. Each layer corresponds to a \synt{layer} definition from
  Figure~\ref{fig:immix-layout}. Each layer is then further demarcated into \synt{field}
  fields, \synt{demarc-val} values, and other \synt{layer} layers.}
\end{figure}

\begin{figure}
\def \mintFontSize {\fontsize{8}{9.5}}
\begin{minted}[xleftmargin=1.7em,linenos=true,fontsize=\mintFontSize,breaklines=true]{\ourlangL}
Region<num_blocks, lines, wrds> -> seq {
  Space @(2^19 bytes)@ -> union {
      num_blocks union {
          FreeBlock @(2^16 bytes)@ -> seq {
              2^16 bytes }
        | Block ||2^16 bytes|| @(2^16 bytes)@
              contains(Line) -> seq {
              cells : # union {
                  FreeCell @(1 words)@ -> # words
                | Cell },
              remainder : # words,
              limit     : 0 words } }
    | wrds (1 words)
    | lines Line @|2^8 bytes|@
          contains(Cell) -> # bytes },
  lms   : lines LineMark,
  refs  : wrds RefBits,
  mks   : wrds MarkBits }
Cell @(1 words)@ contains(Word) -> union {
  seq { cell_0   : Cell ptr,
        cell_1   : Cell ptr, cell_2 : Cell ptr,
        cell_3   : Cell ptr, payload : # words }
  | # words }
RefBits ||1 bytes|| -> bits {
    SHORT_ENCODE  : 1 bits,
    OBJ_START     : 1 bits,
    REF           : 6 bits }
LineMark -> enum { Free | Live | FreshAlloc
                 | ConservLive | PrevLive }
MarkBits ||1 bytes|| -> bits { MARK : 8 bits }
Stk -> seq { stack : # seq { Cell ptr },
             lowWater : 0 words }
Registers -> seq { regs : # seq { Cell ptr },
                   regsEnd : 0 words }
\end{minted}
\def \mintFontSize {\small}
\caption{\label{fig:immix-layout} The \ourlang~specification of immix as implemented in Rust~\cite{rust-gc}.}
\end{figure}

In this section, Figure~\ref{fig:demarc-region} introduces the notion of a demarcation diagram
and Figure~\ref{fig:immix-layout} shows the \ourlang~specification of immix in Rust.
For a precise handling
of \ourlang~semantics, see Section~\ref{sec:semantics}. Throughout this
section subscripts on words$_{\bf 1}$ indicate line numbers in
Figure~\ref{fig:immix-layout}.

\subsection{Immix specification} \label{sec:immix-specification}

Figure~\ref{fig:immix-layout} shows the \ourlang~specification for the Rust
implementation~\cite{rust-gc} of the immix garbage collection algorithm.
The heap is represented as a Region$_{\bf 1}$ parametrized by three formal
arguments: the number of blocks, lines, and
number of words \texttt{wrds} in the region. Note that once \texttt{num_blocks} is
fixed, the other two take on fixed values.\footnote{The default immix
heap is half a gigabyte of memory: $8000$ blocks,
more than $2$ million lines, and over $65$ million words on a $64$-bit machine.}
This constraint we have made is
self-imposed, and not a part of Floorplan semantics. We debated including a version
of the \synt{union} operator which enforces size-equivalence of constituents,
but decided against it for simplicity reasons: two flavors of the union operator arguably
degrades comprehensibility.

A Region$_{\bf_1}$ layer consists of a single Space$_{\bf 2}$ followed by some
metadata fields for marking lines$_{\bf 16}$, looking up reference bytes$_{\bf
17}$, and setting mark bits$_{\bf 18}$. RefBits$_{\bf 24}$ and MarkBits$_{\bf
30}$ both represent bit-fields which consume one byte of memory. Note
that bit order for a RefBits$_{\bf 24}$ is defined such that the
\texttt{OBJ_START}$_{\bf 26}$ bit occurs at a less significant bit than the
\texttt{REF}$_{\bf 27}$ bits. \texttt{SHORT_ENCODE}$_{\bf 25}$ is the least
significant (ones) bit.


Notice here$_{\bf 7}$ that a block of memory is annotated with
the fact that it contains lines. The annotation indicates to the
\ourlang~compiler that it should generate code for converting between a
Line$_{\bf 14}$ and its containing Block$_{\bf 6}$, and vice-versa. The
conditions under which this code gets generated relies on the presence of known
sizes and alignments for lines and blocks respectively.


In this~\cite{rust-gc} version of
immix, objects do not have a header word. Instead each cell's corresponding
RefBits$_{\bf 24}$ in the refs$_{\bf 17}$ array tracks which words$_{\bf 13}$
of memory in the Space$_{\bf 2}$ correspond to the start of an object,
\texttt{OBJ_START}$_{\bf 26}$. The implementation of the immix algorithm
determines how many heap references are in some cell by looking up the first
$4$ bits of the corresponding \texttt{REF}$_{\bf 27}$ field of that cell's
RefBits$_{\bf 17}$.

This implementation of immix extracts the application's root set directly
from the stack and registers. The Stk$_{\bf
31}$ and Registers$_{\bf 33}$ are both assumed to be some number of Cell
pointers$_{\bf 31,33}$ followed by a lowWater$_{\bf 32}$ mark and regsEnd$_{\bf
34}$ ending address respectively. The implementation performs conservative
stack and register scanning.





\section{Semantics} \label{sec:semantics}

\newlength{\cfactor}

\begin{figure}
\begin{minipage}{\linewidth}
\centering
{\bf Core value syntax}\\\smallskip
\begin{tabular}{llcl}\toprule[1pt]
Address     & $\alpha$ & $\in$ & $\mathbb{N}$ \\
Identifier  & $\ell,f$ & $\in$ & Strings \\
Values      & $\nu$ & ::= & $1$ bytes | $0$ bytes | T $\nu_1$ $\nu_2$ | N $\ell$ $\nu$ \\
Type        & $\tau$ & ::= & \{ $\nu$ \} \\
\bottomrule[0.75pt]
\end{tabular}\par
\end{minipage}
\caption{\label{fig:core-value-syntax} Syntactic forms of core \ourlang~values. $\ell$ is for
  \synt{layer-id} and \synt{field-id}, while $f$ is for a \synt{formal-id}.}
\end{figure}

\begin{figure*}
\begin{minipage}{.3\linewidth}
\begin{center}
\begin{tikzpicture}[scale=0.1]
\tikzstyle{every node}+=[inner sep=0pt]
\draw [black] (16.3,-20.3) circle (3);
\draw (16.3,-20.3) node {hd};
\draw [black] (34.2,-20.3) circle (3);
\draw (34.2,-20.3) node {tl};
\draw [black] (34.2,-9.3) circle (3);
\draw (34.2,-9.3) node {$\texttt{K}_0$};
\draw [black] (16.3,-31.5) circle (3);
\draw (16.3,-31.5) node {$1$};
\draw [black] (16.3,-31.5) circle (2.4);
\draw [black] (25.9,-31.5) circle (3);
\draw (25.9,-31.5) node {lft};
\draw [black] (34.2,-31.5) circle (3);
\draw (34.2,-31.5) node {rgt};
\draw [black] (25.9,-46) circle (3);
\draw (25.9,-46) node {$1$};
\draw [black] (25.9,-46) circle (2.4);
\draw [black] (34.2,-46) circle (3);
\draw (34.2,-46) node {$1$};
\draw [black] (34.2,-46) circle (2.4);
\draw [black] (42.8,-46) circle (3);
\draw (42.8,-46) node {$1$};
\draw [black] (42.8,-46) circle (2.4);
\draw [black] (51.6,-45.9) circle (3);
\draw (51.6,-45.9) node {$1$};
\draw [black] (51.6,-45.9) circle (2.4);
\draw [black] (31.64,-10.87) -- (18.86,-18.73);
\fill [black] (18.86,-18.73) -- (19.8,-18.74) -- (19.28,-17.88);
\draw [black] (34.2,-12.3) -- (34.2,-17.3);
\fill [black] (34.2,-17.3) -- (34.7,-16.5) -- (33.7,-16.5);
\draw [black] (16.3,-23.3) -- (16.3,-28.5);
\fill [black] (16.3,-28.5) -- (16.8,-27.7) -- (15.8,-27.7);
\draw [black] (32.41,-22.71) -- (27.69,-29.09);
\fill [black] (27.69,-29.09) -- (28.56,-28.74) -- (27.76,-28.15);
\draw [black] (34.2,-23.3) -- (34.2,-28.5);
\fill [black] (34.2,-28.5) -- (34.7,-27.7) -- (33.7,-27.7);
\draw [black] (25.9,-34.5) -- (25.9,-43);
\fill [black] (25.9,-43) -- (26.4,-42.2) -- (25.4,-42.2);
\draw [black] (34.2,-34.5) -- (34.2,-43);
\fill [black] (34.2,-43) -- (34.7,-42.2) -- (33.7,-42.2);
\draw [black] (35.73,-34.08) -- (41.27,-43.42);
\fill [black] (41.27,-43.42) -- (41.29,-42.48) -- (40.43,-42.99);
\draw [black] (36.51,-33.41) -- (49.29,-43.99);
\fill [black] (49.29,-43.99) -- (48.99,-43.09) -- (48.35,-43.86);
\end{tikzpicture}
\end{center}

  \rule{\linewidth}{0.75pt}
  \begin{center}
  $\texttt{K}_0$: $n = 1$, $\#_0 = 3$ bytes
\end{center}
\end{minipage}
\begin{minipage}{.3\linewidth}
\begin{center}
\begin{tikzpicture}[scale=0.1]
\tikzstyle{every node}+=[inner sep=0pt]
\draw [black] (16.3,-20.3) circle (3);
\draw (16.3,-20.3) node {hd};
\draw [black] (34.2,-20.3) circle (3);
\draw (34.2,-20.3) node {tl};
\draw [black] (34.2,-9.3) circle (3);
\draw (34.2,-9.3) node {$\texttt{K}_1$};
\draw [black] (16.3,-31.5) circle (3);
\draw (16.3,-31.5) node {$1$};
\draw [black] (16.3,-31.5) circle (2.4);
\draw [black] (25.9,-31.5) circle (3);
\draw (25.9,-31.5) node {lft};
\draw [black] (34.2,-31.5) circle (3);
\draw (34.2,-31.5) node {rgt};
\draw [black] (25.9,-46) circle (3);
\draw (25.9,-46) node {$1$};
\draw [black] (25.9,-46) circle (2.4);
\draw [black] (43,-46) circle (3);
\draw (43,-46) node {$1$};
\draw [black] (43,-46) circle (2.4);
\draw [black] (43,-31.5) circle (3);
\draw (43,-31.5) node {lft};
\draw [black] (51.7,-31.5) circle (3);
\draw (51.7,-31.5) node {rgt};
\draw [black] (9.7,-31.5) circle (3);
\draw (9.7,-31.5) node {$1$};
\draw [black] (9.7,-31.5) circle (2.4);
\draw [black] (51.7,-46) circle (3);
\draw (51.7,-46) node {$1$};
\draw [black] (51.7,-46) circle (2.4);
\draw [black] (31.64,-10.87) -- (18.86,-18.73);
\fill [black] (18.86,-18.73) -- (19.8,-18.74) -- (19.28,-17.88);
\draw [black] (34.2,-12.3) -- (34.2,-17.3);
\fill [black] (34.2,-17.3) -- (34.7,-16.5) -- (33.7,-16.5);
\draw [black] (16.3,-23.3) -- (16.3,-28.5);
\fill [black] (16.3,-28.5) -- (16.8,-27.7) -- (15.8,-27.7);
\draw [black] (32.41,-22.71) -- (27.69,-29.09);
\fill [black] (27.69,-29.09) -- (28.56,-28.74) -- (27.76,-28.15);
\draw [black] (34.2,-23.3) -- (34.2,-28.5);
\fill [black] (34.2,-28.5) -- (34.7,-27.7) -- (33.7,-27.7);
\draw [black] (25.9,-34.5) -- (25.9,-43);
\fill [black] (25.9,-43) -- (26.4,-42.2) -- (25.4,-42.2);
\draw [black] (36.05,-22.66) -- (41.15,-29.14);
\fill [black] (41.15,-29.14) -- (41.05,-28.2) -- (40.26,-28.82);
\draw [black] (36.73,-21.92) -- (49.17,-29.88);
\fill [black] (49.17,-29.88) -- (48.77,-29.03) -- (48.23,-29.87);
\draw [black] (43,-34.5) -- (43,-43);
\fill [black] (43,-43) -- (43.5,-42.2) -- (42.5,-42.2);
\draw [black] (14.78,-22.88) -- (11.22,-28.92);
\fill [black] (11.22,-28.92) -- (12.06,-28.48) -- (11.2,-27.97);
\draw [black] (51.7,-34.5) -- (51.7,-43);
\fill [black] (51.7,-43) -- (52.2,-42.2) -- (51.2,-42.2);
\end{tikzpicture}
\end{center}

  \rule{\linewidth}{0.75pt}
  \begin{center}
  $\texttt{K}_1$: $n = 2$, $\#_0 = 0$ bytes, $\#_1 = 1$ byte
\end{center}
\end{minipage}
\begin{minipage}{.3\linewidth}
\begin{center}
\begin{tikzpicture}[scale=0.1]
\tikzstyle{every node}+=[inner sep=0pt]
\draw [black] (16.3,-20.3) circle (3);
\draw (16.3,-20.3) node {hd};
\draw [black] (34.2,-20.3) circle (3);
\draw (34.2,-20.3) node {tl};
\draw [black] (34.2,-9.3) circle (3);
\draw (34.2,-9.3) node {$\texttt{K}_2$};
\draw [black] (16.3,-31.5) circle (3);
\draw (16.3,-31.5) node {$1$};
\draw [black] (16.3,-31.5) circle (2.4);
\draw [black] (25.9,-31.5) circle (3);
\draw (25.9,-31.5) node {lft};
\draw [black] (34.2,-31.5) circle (3);
\draw (34.2,-31.5) node {rgt};
\draw [black] (25.9,-46) circle (3);
\draw (25.9,-46) node {$1$};
\draw [black] (25.9,-46) circle (2.4);
\draw [black] (43,-46) circle (3);
\draw (43,-46) node {$1$};
\draw [black] (43,-46) circle (2.4);
\draw [black] (43,-31.5) circle (3);
\draw (43,-31.5) node {lft};
\draw [black] (51.7,-31.5) circle (3);
\draw (51.7,-31.5) node {rgt};
\draw [black] (9.7,-31.5) circle (3);
\draw (9.7,-31.5) node {$1$};
\draw [black] (9.7,-31.5) circle (2.4);
\draw [black] (34.2,-46) circle (3);
\draw (34.2,-46) node {$1$};
\draw [black] (34.2,-46) circle (2.4);
\draw [black] (31.64,-10.87) -- (18.86,-18.73);
\fill [black] (18.86,-18.73) -- (19.8,-18.74) -- (19.28,-17.88);
\draw [black] (34.2,-12.3) -- (34.2,-17.3);
\fill [black] (34.2,-17.3) -- (34.7,-16.5) -- (33.7,-16.5);
\draw [black] (16.3,-23.3) -- (16.3,-28.5);
\fill [black] (16.3,-28.5) -- (16.8,-27.7) -- (15.8,-27.7);
\draw [black] (32.41,-22.71) -- (27.69,-29.09);
\fill [black] (27.69,-29.09) -- (28.56,-28.74) -- (27.76,-28.15);
\draw [black] (34.2,-23.3) -- (34.2,-28.5);
\fill [black] (34.2,-28.5) -- (34.7,-27.7) -- (33.7,-27.7);
\draw [black] (25.9,-34.5) -- (25.9,-43);
\fill [black] (25.9,-43) -- (26.4,-42.2) -- (25.4,-42.2);
\draw [black] (36.05,-22.66) -- (41.15,-29.14);
\fill [black] (41.15,-29.14) -- (41.05,-28.2) -- (40.26,-28.82);
\draw [black] (36.73,-21.92) -- (49.17,-29.88);
\fill [black] (49.17,-29.88) -- (48.77,-29.03) -- (48.23,-29.87);
\draw [black] (43,-34.5) -- (43,-43);
\fill [black] (43,-43) -- (43.5,-42.2) -- (42.5,-42.2);
\draw [black] (14.78,-22.88) -- (11.22,-28.92);
\fill [black] (11.22,-28.92) -- (12.06,-28.48) -- (11.2,-27.97);
\draw [black] (34.2,-34.5) -- (34.2,-43);
\fill [black] (34.2,-43) -- (34.7,-42.2) -- (33.7,-42.2);
\end{tikzpicture}
\end{center}

  \rule{\linewidth}{0.75pt}
  \begin{center}
  $\texttt{K}_2$: $n = 2$, $\#_0 = 1$ byte, $\#_1 = 0$ bytes
\end{center}
\end{minipage}

\caption{\label{fig:fsm} The three layouts for Code~F\ref{code:core-calculus}, with satisfying assignments
to Equation~\thesection.\ref{eqn:reduction-example}.}
\end{figure*}
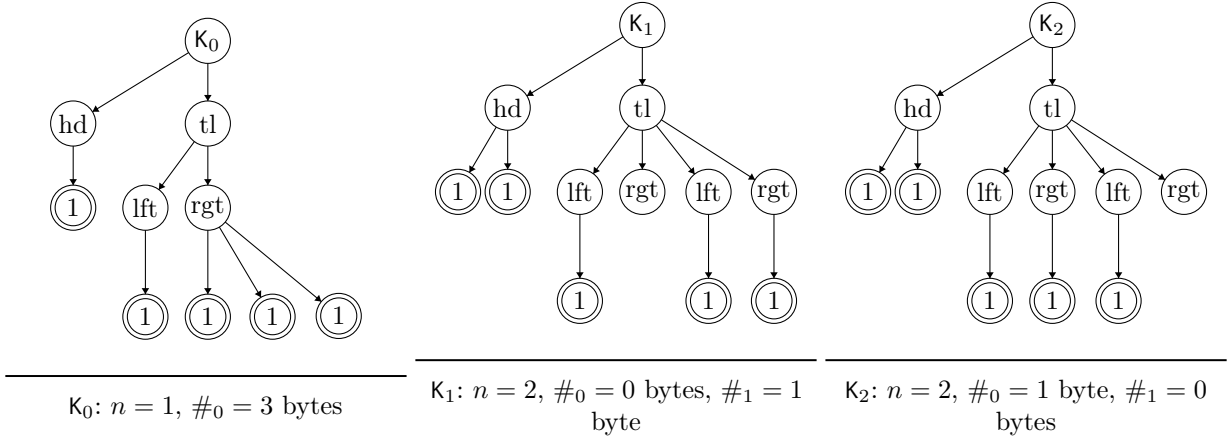

\begin{figure}
\begin{minipage}{\linewidth}
\centering
{\bf Core calculus}\\\smallskip
\begin{tabular}{llcl}\toprule[1pt]
Nats & $n, m, c$ & $\in$ & $\mathbb{N}$ \\
Alignment & $\hat{a}$ & $\in$ & $\mathbb{N}^+$ \\
  Exp & $e$ & ::= & Prim $n$ | $\Con{n}{e}$ | $e$ @ $\hat{a}$ \\
 & & | & $e_1 + e_2$ | $e_1 \OR e_2$ | $y \HasType e$ | $\Exists{f}{e}$ \\
 & & | & $f \Pound e$ \\
Size & $\delta$ & ::= & $m$ \\
Environment & $\theta$ & ::= & $\{ f \mapsto n \}$ \\
Config & $\chi$ & ::= & $(\alpha, \delta, e)$ \\
\bottomrule[0.75pt]
\end{tabular}\par
\end{minipage}
\caption{\label{fig:core-calculus} Core expression language representing a
\ourlang~specification. A function $\bar{\gamma}$ with type $\chi
\rightarrow \tau$ models the semantics of a memory layout.}
\end{figure}

In this section we present the semantics of a \ourlang~specification to be the
set of heaps which satisfy the specification, with satisfaction as
defined in Figure~\ref{fig:spec-semantics}. Note that Floorplan semantics do not
suffice to ingest raw pages.


%

\subsection{Concrete value semantics}


We represent an instance of a memory layout as a tree, as in
Figure~\ref{fig:core-value-syntax}. Addresses are natural numbers representing
locations in a flat addressable sequence of bytes. A value is a rooted binary
tree with leaves each representing either zero or one byte. Trees may be
interspersed with named ``N'' components,
mapping directly back to named types in a \ourlang~specification, as will
become apparent by the semantics in the following
Section~\ref{sec:spec-semantics}. A finite set of trees represents a concrete
type of memory.

An in-order traversal over a tree defines the order in which bytes at
the leaves of the tree occur contiguously in memory.
Finally, $\textrm{leaves}(\nu)$ computes the number of 1-byte leaves in the tree
as defined below, equivalent to the number of bytes
the tree consumes in memory.

\smallskip
\begin{center}
\begin{tabular}{lcl}\toprule[1pt]
$\textrm{leaves}(\textrm{1 bytes})$ & $=$ & 1 \\
$\textrm{leaves}(\textrm{0 bytes})$ & $=$ & 0 \\
$\textrm{leaves}(T$ $\nu_1$ $\nu_2)$ & $=$ & $\textrm{leaves}(\nu_1) + \textrm{leaves}(\nu_2)$ \\
$\textrm{leaves}(N$ $\ell$ $\nu)$ & $=$ & $\textrm{leaves}(\nu)$ \\
\bottomrule[0.75pt]
\end{tabular}
\end{center}


%
%
%
%

\subsubsection{Example: the trees of a specification}

Before introducing the core calculus, take the following Floorplan declaration:

\begin{tabular}{p{0.80\columnwidth} p{0.20\columnwidth}}
\begin{floorplancode}
  K<n> |5 bytes| -> seq { hd : n (1 bytes),
    tl : n seq { lft : 1 bytes, rgt : # bytes } }
\end{floorplancode}
  & ~\newCodeC{code:core-calculus}
\end{tabular}

This code represents the three distinct memory layouts as depicted
in Figure~\ref{fig:fsm}, one for each feasible assignment of natural numbers
to $n$ and the ``\texttt{\#}''. The $n = 0$ case is not feasible because that
case consumes $0 \neq 5$ bytes. Similarly the $n = 3$ case is not feasible because
the \texttt{hd} consumes $3$ bytes and the \texttt{tl} consumes \emph{at least}
$3$ bytes, one for each copy of \texttt{lft}, which sums to at least $6 \neq 5$ bytes.
Formally, for constants $n, \#_i \in \mathbb{N}$, memory
layout instances must satisfy the following constraint
satisfaction~\cite{ConstraintBook} equation:

\noindent
\begin{tabular}{p{0.80\columnwidth} p{0.20\columnwidth}}
$$n + \sum_{i=0}^{n-1} n \ast (1 + \#_i) = 5$$
& ~\linebreak \newEqn{eqn:reduction-example}
\end{tabular}

Equation~\refEqn{eqn:reduction-example} above was written by hand, and is not
formally synthesized by the compiler. We will see in Section~\ref{sec:spec-semantics-example}
how to reduce Code~F\ref{code:core-calculus} to tree $\texttt{K}_0$ from Figure~\ref{fig:fsm}.

\subsection{Abstract expression semantics\protect\footnotemark} \label{sec:spec-semantics}

\footnotetext{Subscripts$_{\bf 1}$ on$_{\bf 2}$ words$_{\bf 3}$%
 correspond to lines in Figure~\ref{fig:spec-semantics}}


Now 
we define the core expression language as in Figure~\ref{fig:core-calculus}.
Each expression $e$ denotes a memory layout. A memory layout has a
corresponding (possibly empty) set of values $\nu$ representing a type $\tau$
computable by the memory layout modeling function $\gamma$ in
Figure~\ref{fig:spec-semantics}. A primitive expression (Prim $n$) denotes a
contiguous (possibly empty$_{\bf 7}$) sequence$_{\bf 9}$ of $n$ bytes.
Similarly, a constrained expression ($\Con{n}{e}$) denotes a contiguous sequence
of $n$ bytes, but only for the (possibly non-existent$_{\bf 13}$) memory layout
instances for which the substructure denoted by $e$ fits precisely$_{\bf 12}$
into $n$ bytes. An aligned expression ($e$ @ $\hat{a}$) denotes a memory layout
for which the address of the first byte of memory of the layout must be a
natural number multiple$_{\bf 15}$ of $\hat{a}$ bytes.

The remaining operators are the concatenation ``$+$'' and
union ``$\OR$'' binary operators, as well as name binding with $y \HasType e$. A
concatenation of two expressions denotes the contiguously laid out sequence$_{2-4}$ of
those two expressions. A union of two expressions denotes a left-most aligned
instance of either the first$_{\bf 5}$ or the second$_{\bf 6}$ expression.
A named expression $y \HasType e$ binds$_{\bf 17,18}$
the name $y$ to the expression $e$. An existentially quantified expression
$\Exists{f}{e}$ brings the variable $f$ into scope$_{\bf 20}$ in the subexpression $e$.

A variable on a repetition, the $f$ in $(f \Pound e)$, may be referenced multiple times.
Each reference must also
take on the same fixed value. This feature causes a \ourlang~specification (i.e. grammar)
to be non-regular: there exist \ourlang~grammars which fail the Pumping Lemma.

\setlength{\cfactor}{.9cm}
\begin{figure}
\begin{minipage}{\linewidth}
\centering
{\bf Memory layout model $\bar{\gamma}$}\\\smallskip
\begin{tabular}{llcl}\toprule[1pt]
  1 & $\gamma \lbparen (\alpha, m, \theta, e_1 + e_2) \rbparen$ & & \\
  2 & & & \hspace{-4.5\cfactor} $=$ \{ $T$ $r_1$ $r_2$ $\mid$ $r_1 \in \bigcup\limits_{i=0}^{m} \gamma \lbparen(\alpha, i, \theta, e_1)\rbparen$ \\
  3 & & & \hspace{-2.8\cfactor} , $r_2 \in \gamma\lbparen(\alpha + \textrm{leaves}(r_1)$ \\
  4 & & & \hspace{-2.0\cfactor} $, m - \textrm{leaves}(r_1), \theta, e_2)\rbparen\}$ \\
  5 & $\gamma \lbparen (\alpha, m, \theta, e_1 \OR e_2) \rbparen$ & $=$ &
  \hspace{0.3\cfactor} $\gamma\lbparen(\alpha, m, \theta, e_1)\rbparen$ \\
  6 & & & $\bigcup \gamma\lbparen(\alpha, m, \theta, e_2)\rbparen$ \\
  7 & $\gamma \lbparen (\alpha, 0, \theta, \textrm{Prim } 0) \rbparen$ & $=$ & \{ 0 bytes \} \\
  8 & $\gamma \lbparen (\alpha, m, \theta, \textrm{Prim } n) \rbparen$ & & \\
  9 & & & \hspace{-4.5\cfactor} | $m \equiv n =$ \{ $T$ (1 bytes$)_1$ (~$\cdots$ T (1 bytes$)_n$ (0 bytes)) \} \\
  10 & & & \hspace{-4.5\cfactor} | $m \neq n = \emptyset$ \\
  11 & $\gamma \lbparen (\alpha, m, \theta, \Con{n}{e}) \rbparen$ & & \\
  12 & & & \hspace{-4.5\cfactor} | $m \equiv n = \gamma\lbparen(\alpha,m,\theta,e)\rbparen$ \\
  13 & & & \hspace{-4.5\cfactor} | $m \neq n = \emptyset$ \\
  14 & $\gamma \lbparen (\alpha, m, \theta, \textrm{e @ } \hat{a}) \rbparen$ & & \\
  15 & & & \hspace{-4.5\cfactor} | $\alpha \bmod \hat{a} \equiv 0 = \gamma\lbparen(\alpha,m,\theta,e)\rbparen$ \\
  16 & & & \hspace{-4.5\cfactor} | $\alpha \bmod \hat{a} \neq 0 = \emptyset$ \\
  17 & $\gamma \lbparen (\alpha, m, \theta, \ell \HasType e) \rbparen$ & \hspace{-1\cfactor} $=$ & \hspace{-.8\cfactor} \{ $N$ $\ell$ $r$ \\
  18 & & & \hspace{-.8\cfactor} $\mid$ $r \in \gamma \lbparen (\alpha, m, \theta, e) \rbparen$ \} \\
  19 & $\gamma \lbparen (\alpha, m, \theta, \Exists{f}{e}) \rbparen$ & \hspace{-1\cfactor} $=$ & \\
  20 & & & \hspace{-3.5\cfactor} $\bigcup\limits_{i=0}^{m} \gamma \lbparen (\alpha, m, \theta \{ f \mapsto i \}, e) \rbparen$ \\


  21 & $\gamma \lbparen (\alpha,m,\theta, f \Pound e) \rbparen$ & & \\
  22 & & & \hspace{-4.5\cfactor} | $f \notin \textrm{dom}(\theta) = \emptyset$ \\
  23 & & & \hspace{-4.5\cfactor} | $m \equiv \theta(f) \equiv 0 =$ \{ $T$ (0 bytes) (0 bytes) \} \\
  24 & & & \hspace{-4.5\cfactor} | $\theta(f) \equiv 0 = \emptyset$ \\
  25 & & & \hspace{-4.5\cfactor} | $\theta(f) > 0$ \\

  26 & & & \hspace{-4.5\cfactor} $=$ \{ $T$ $r_1$ $r_2$ \\
  27 & & & \hspace{-4.15\cfactor} $\mid$ $r_1 \in \bigcup\limits_{i=0}^{m} \gamma \lbparen (\alpha, i, \theta, e) \rbparen$ \\
  28 & & & \hspace{-4.15\cfactor} , $r_2 \in \gamma \lbparen (\alpha + \textrm{leaves}(r_1), m - \textrm{leaves}(r_1)$ \\
  29 & & & \hspace{-3.0\cfactor}  $, \theta \{f \mapsto (\theta(f) - 1) \}, f \Pound e)\rbparen$ \\
  30 & & & \hspace{-4.15\cfactor} , $m \equiv \textrm{leaves}(r_1) + \textrm{leaves}(r_2)$ \} \\
  31 & $\bar{\gamma} \lbparen(\alpha, m, e)\rbparen$ & $=$ & $\gamma \lbparen (\alpha, m, \emptyset, e) \rbparen$ \\
\bottomrule[0.75pt]
\end{tabular}\par
\end{minipage}
\caption{\label{fig:spec-semantics} Denotational semantics of \ourlang.}
\end{figure}


\subsubsection{Denotations: reducing Code~F\ref{code:core-calculus} to core values} \label{sec:spec-semantics-example}

Figure~\ref{fig:spec-semantics} shows our core denotational semantics,
the first three parameters of which are $\gamma$: $\alpha$, $m$, and $\theta$.
$\alpha$ represents the base address of a memory layout, $m$ represents the precise number of
bytes in which the layout must fit, and $\theta$ represents a name environment.
As the compilation rules (upcoming in Section~\ref{sec:compilation-rules}) are not
particularly important to understand core Floorplan semantics, we give
Code~F\ref{code:core-calculus} translated to the core calculus here:

\refstepcounter{codeCore}
\refstepcounter{codeCore}%
\refstepcounter{codeCore}%
\refstepcounter{codeCore}%
\refstepcounter{codeCore}%


\noindent
\begin{minipage}[t]{0.9\columnwidth}
{\begin{align*}
  \hspace{0.5cm} &  \texttt{K} \HasType ( \exists~n~.~\textrm{Con } 5~( & \\
  \hspace{0.5cm} & \hspace{0.6cm}                      (\texttt{hd} \HasType n~\Pound~(\textrm{Prim 1}))~+ & \\
  \hspace{0.5cm} & \hspace{0.6cm}                        (\texttt{tl} \HasType n~\Pound~((\texttt{lft} \HasType (\textrm{Prim~} 1)) & \\
  \hspace{0.5cm} & \hspace{0.6cm}                       \textrm{\hspace{0.5cm}} + (\texttt{rgt} \HasType (\exists~f_0~.~f_0~\Pound~(\textrm{Prim 1})))))) & 
\end{align*}}
\end{minipage}%
\begin{minipage}[t]{0.09\columnwidth}
~\linebreak\linebreak\linebreak (C\refstepcounter{codeCore}\label{code:core:core-calculus}\arabic{codeCore})
\end{minipage}

Of note on the fourth line of C\ref{code:core:core-calculus}, the existentially bound $f_0$ variable
materializes by way of Rule (2) of Figure~\ref{fig:compilation-rules}.
Furthermore, listed below are the steps through the semantics in
that figure for reducing Code C\ref{code:core:core-calculus} to the \texttt{hd} sub-branch of the
left-most tree $\texttt{K}_0$ of Figure~\ref{fig:fsm}:


{\small
\newcommand{\tikzmark}[1]{\tikz[baseline=-0.5ex, overlay, remember picture] \coordinate (#1);}
\smallskip%
\noindent%
{\hspace{0.8mm} \begin{tabular}{@{}l|l|p{2.8cm}|l}
  {\bf Line} & {\bf Exp} & {\bf Trees} & {\bf Step} \\\hline
  {\small 17,18} & \texttt{K} $\HasType$ $e_1$ & $\gamma \lbparen 0,5,\emptyset,e_1 \rbparen$ & Pick $m=5$ \\
  {\small 19,20} & $e_1 = \exists~n~.~e_2$ & Let $\theta_1=\{n \mapsto 1\}$ in $\gamma \lbparen 0,5,\theta_1,e_2 \rbparen$ & Pick $i=1$ \\
  {\small 11,12} & $e_2 = $ Con $5~e_3$ & $\gamma\lbparen 0,5,\theta_1,e_3\rbparen$ & Reduce Con \\
  {\small 1,2} & $e_3 = e_{\texttt{hd}} + e_{\texttt{tl}}$ & $\gamma\lbparen 0,1,\theta_1,e_{\texttt{hd}}\rbparen$ & Pick $i=1$ %
  \hspace{1.6mm} \tikzmark{plus1}\\
  {\small 17,18} & $e_{\texttt{hd}} = $ hd $\HasType e_4$ & $\gamma\lbparen0,1,\theta_1,e_4\rbparen$ & Reduce name \\
  {\small 21,27} & $e_4 = n \Pound e_5$ & $\gamma\lbparen 0,1,\theta_1,e_5\rbparen$ & Pick $i=1$%
  \hspace{2.0mm} \tikzmark{pound1} \\
  {\small 8,9} & $e_5 = $ Prim $1$ & T ($1$ bytes) ($0$ bytes) & Eval tree \\
  {\small 3,4} & $e_4 = n \Pound e_5$ & $\gamma \lbparen 0+1,1-1,\theta_1,e_4 \rbparen$ & Resume $(\Pound)$ \tikzmark{pound2} \\
  {\small 23} & $e_4 = n \Pound e_5$ & T ($0$ bytes) ($0$ bytes) & Eval tree \\
  {\small 3,4} & $e_{\texttt{hd}} + e_{\texttt{tl}}$ & $\gamma\lbparen 0+1,5-1,\theta_1,e_{\texttt{tl}}\rbparen$ & Resume $(+)$ \tikzmark{plus2} \\
  {\small 17,18} & $e_{\texttt{tl}} = $ tl $\HasType e_6$ & $\gamma\lbparen 1,4,\theta_1,e_6\rbparen$ & Skip $e_{\texttt{tl}}$ \\
\end{tabular}}%
\smallskip
\tikz[overlay, remember picture]{
  \coordinate (plusA) at ($ (plus1) + (0.3,0) $);
  \coordinate (plusB) at ($ (plus2) + (0.3,0) $);
  \coordinate (poundA) at ($ (pound1) + (0.2,0) $);
  \coordinate (poundB) at ($ (pound2) + (0.2,0) $);
  \draw [] (plus1) -- (plusA);
  \draw [] (plusA) -- (plusB);
  \draw [->] (plusB) -- (plus2);
  \draw [] (pound1) -- (poundA);
  \draw [] (poundA) -- (poundB);
  \draw [->] (poundB) -- (pound2);
}
}
\smallskip

We manually picked values for $m$ and $i$ in order to
derive tree $\texttt{K}_0$. These derivation steps compute the first line of
the following tree in data type form, with \texttt{B0} and \texttt{B1} representing
$0$ bytes and $1$ bytes respectively:

\smallskip
\begin{minipage}[t]{0.95\columnwidth}
\begin{verbatim}N "K" (T (N "hd" (T (T B1 B0) (T B0 B0)))
  (N "tl" (T (T (N "lft" (T B1 B0)) 
                (N "rgt" (T (T B1 B0)
      (T (T B1 B0) (T (T B1 B0) (T B0 B0)))))
  ) (T B0 B0))))
\end{verbatim}
\end{minipage}
\smallskip



Certain properties of the denotational semantics from Figure~\ref{fig:spec-semantics}
have been proved\footnote{Available in extended version.} correct in Coq. Such properties include that $\gamma$ always returns
trees with $m$ one-byte leaves and that $\gamma$ is a total computable function.

\subsection{Compilation rules} \label{sec:compilation-rules}

\def\ruleNumberOffset{-.55}
\setlength{\cfactor}{1.05cm}
\begin{figure}
{\small
\begin{tabular}{llcl}
  & $\mathbb{C} \lbparen \synt{layer-simple} \rbparen =$ & & \\
  & $\mathbb{C} \lbparen$ \synt{layer-id} (\texttt{`\textless'} \synt{formals} \texttt{`\textgreater'})? \synt{mag}? \synt{align}? \texttt{`-\textgreater'} \\
  & \hspace{.6\cfactor}\synt{demarc-val} $\rbparen$, $f_i \in \synt{formals}$ & & \\
  & \hspace{\ruleNumberOffset\cfactor} (1) & & \hspace{-8\cfactor} $\vDash \synt{layer-id} \HasType (\Exists{f_0}{\cdots \Exists{f_n}$ \\
  & & & \hspace{-7.7\cfactor} $(\mathbb{M}\lbparen \synt{mag} \rbparen}$ \\
  & & & \hspace{-7.4\cfactor} $(\mathbb{C}\lbparen \synt{demarc-val} \rbparen
    \Align (\Delta_{\textrm{byte}} \lbparen \synt{align} \rbparen))))$ \\

  & \rule{.5\linewidth}{.5pt} & & \\
  & $\mathbb{C} \lbparen \synt{demarc-val} \rbparen =$
  $\mathbb{C} \lbparen$ \texttt{`\#'} \synt{demarc-val} $\rbparen$ & & \\
  & \hspace{\ruleNumberOffset\cfactor} (2) & & \hspace{-8\cfactor} $\vDash$ let $f$ = \texttt{fresh}(\synt{demarc-val}) \\
  & & & \hspace{-7.7\cfactor}        in~~$\Exists{f}{f \Pound \mathbb{C}\lbparen \synt{demarc-val} \rbparen}$ \\

  & \rule{.5\linewidth}{.5pt} & & \\
  & $\mathbb{C} \lbparen \synt{demarc-val} \rbparen =$
  $\mathbb{C} \lbparen$ \synt{formal-id} \synt{demarc-val} $\rbparen$ & & \\
  & \hspace{\ruleNumberOffset\cfactor} (3) & & \hspace{-8\cfactor} $\vDash \synt{formal-id} \Pound \mathbb{C}\lbparen \synt{demarc-val} \rbparen$ \\

  & \rule{.5\linewidth}{.5pt} & & \\
  & $\mathbb{C} \lbparen \synt{seq} \rbparen =$
  $\mathbb{C} \lbparen$ \texttt{`seq'} \texttt{`\{'} $\synt{demarc}_0 \cdots \synt{demarc}_n$ \texttt{`\}'} $\rbparen$ \\
  & \hspace{\ruleNumberOffset\cfactor} (4) & & \hspace{-8\cfactor} $\vDash \mathbb{C}\lbparen \synt{demarc}_0 \rbparen + \cdots + \mathbb{C}\lbparen \synt{demarc}_n \rbparen$ \\

  & \rule{.5\linewidth}{.5pt} & & \\
  & $\mathbb{C} \lbparen \synt{union} \rbparen =$ & & \\
  & $\mathbb{C} \lbparen$ \texttt{`union'} \texttt{`\{'} $\synt{demarc}_0 \texttt{`|'} \cdots \texttt{`|'} \synt{demarc}_n \textrm{\texttt{`\}'}} \rbparen$ & & \\
  & \hspace{\ruleNumberOffset\cfactor} (5) & & \hspace{-8\cfactor} $\vDash \mathbb{C} \lbparen \synt{demarc}_0\rbparen \OR \cdots \OR \mathbb{C}\lbparen \synt{demarc}_n\rbparen$ \\

  & \rule{.5\linewidth}{.5pt} & & \\
  & $\mathbb{C} \lbparen \synt{field} \rbparen =$
  $\mathbb{C} \lbparen$ \synt{field-id} \texttt{`:'} \synt{demarc-val} $\rbparen$ \\
  & \hspace{\ruleNumberOffset\cfactor} (6) & & \hspace{-8\cfactor} $\vDash \synt{field-id} \HasType \mathbb{C}\lbparen \synt{demarc-val} \rbparen$ \\

  & \rule{.5\linewidth}{.5pt} & & \\
  & \hspace{\ruleNumberOffset\cfactor} (7) & & \hspace{-8.3\cfactor} $\mathbb{C} \lbparen \synt{ptr} \rbparen =$
  $\mathbb{C} \lbparen$ (\synt{layer-id} | \synt{field-id}) \texttt{\texttt{`ptr'}} $\rbparen$ \\
  & & & \hspace{-8\cfactor} $\vDash \mathbb{C}\lbparen$ 1 word $\rbparen$ \\

  & \rule{.5\linewidth}{.5pt} & & \\
  & $\mathbb{C} \lbparen \synt{enum} \rbparen =$
  $\mathbb{C} \lbparen$ \texttt{`enum'} \texttt{`\{'} $\synt{flag-id}_0 \cdots \synt{flag-id}_n$ \texttt{`\}'} $\rbparen$ \\
  & \hspace{\ruleNumberOffset\cfactor} (8) & & \hspace{-8\cfactor} $\vDash$ Prim $\left\lceil{\log_2(n + 1) * \frac{1 \textrm{ byte}}{8 \textrm{ bits}}}\right\rceil$ \\

  & \rule{.5\linewidth}{.5pt} & & \\
  & $\mathbb{C} \lbparen \synt{bits} \rbparen =$
  $\mathbb{C} \lbparen$ \texttt{`bits'} \texttt{`\{'} $\synt{bits-exp}_0 \cdots \synt{bits-exp}_n$ \texttt{`\}'} $\rbparen$ \\
  & \hspace{\ruleNumberOffset\cfactor} (9) & & \hspace{-8\cfactor} $\vDash$ Prim $\left\lceil \left( \sum\limits_{i = 0}^n (\Delta_{\textrm{bit}}\synt{bits-exp}_i)
                                                     \right) * \frac{\textrm{1 byte}}{\textrm{8 bits}} \right\rceil$ \\
  
  & \rule{.5\linewidth}{.5pt} & & \\
  & \hspace{\ruleNumberOffset\cfactor} (10) & & \hspace{-8.3\cfactor} $\mathbb{C} \lbparen \synt{size-arith} \rbparen \vDash$ Prim $\left(\Delta_{\textrm{byte}} \synt{size-arith}\right)$ \\

\end{tabular}  
  }
\caption[Comp sem]{\label{fig:compilation-rules} Compilation rules for translating
surface syntax to a core expression. Syntax inside oxford-like
brackets\footnotemark~%
is surface syntax, and syntax after a double-turnstile\footnotemark~$\vDash$
is a core expression. Formals support list membership, $\in$.}
\end{figure}

\setlength{\cfactor}{0.2cm}
\begin{figure}
{\footnotesize
\begin{tabular}{l}\toprule[1pt]
  
  $\mathbb{M}\lbparen \texttt{`|'} \synt{size-arith} \texttt{`|'} \rbparen(e)$
    $\vDash \Con{\left(\Delta_{\textrm{bytes}} \lbparen \synt{size-arith} \rbparen\right)}{e}$ \\

  $\Delta_{\textrm{bit}} \equiv \Delta$ \\
  $\Delta_{\textrm{byte}} \equiv \left\lceil \Delta_{\textrm{bit}} * \frac{\textrm{1 byte}}{\textrm{8 bits}} \right\rceil$ \\
  
  $\Delta \lbparen \texttt{`bits'} \rbparen \vDash \frac{\textrm{1 bits}}{\textrm{1 bit}}$, $\Delta \lbparen \texttt{`bytes'} \rbparen \vDash \frac{\textrm{8 bits}}{\textrm{1 byte}}$ \\
  $\Delta \lbparen \texttt{`words'} \rbparen \vDash \frac{c_w\textrm{ bits}}{\textrm{1 word}}$, $\Delta \lbparen \texttt{`pages'} \rbparen \vDash \frac{c_p\textrm{ bits}}{\textrm{1 page}}$ \\

  $\Delta \lbparen \synt{int} \rbparen \vDash \synt{int}$ \\
  $\Delta \lbparen \synt{bin} \rbparen \vDash \texttt{int}(\synt{bin})$ \\
  $\Delta \lbparen \synt{lit-arith}_l \texttt{`+'} \synt{lit-arith}_r \rbparen$
    $\vDash \Delta \lbparen \synt{lit-arith}_l \rbparen + \Delta \lbparen \synt{lit-arith}_r \rbparen $ \\
  $\Delta \lbparen \synt{lit-arith}_l \texttt{`-'} \synt{lit-arith}_r \rbparen$
    $\vDash \Delta \lbparen \synt{lit-arith}_l \rbparen - \Delta \lbparen \synt{lit-arith}_r \rbparen $ \\
  $\Delta \lbparen \synt{lit-arith}_l \texttt{`*'} \synt{lit-arith}_r \rbparen$
    $\vDash \Delta \lbparen \synt{lit-arith}_l \rbparen \ast \Delta \lbparen \synt{lit-arith}_r \rbparen $ \\
  $\Delta \lbparen \synt{lit-arith}_l \texttt{`/'} \synt{lit-arith}_r \rbparen$
    $\vDash \left\lfloor \Delta \lbparen \synt{lit-arith}_l \rbparen / \Delta \lbparen \synt{lit-arith}_r \rbparen \right\rfloor$ \\
  $\Delta \lbparen \synt{lit-arith}_l \texttt{`\^{}'} \synt{lit-arith}_r \rbparen$
    $\vDash \left(\Delta \lbparen \synt{lit-arith}_l \rbparen\right)^{\Delta \lbparen \synt{lit-arith}_r \rbparen}$ \\
  $\Delta \lbparen \synt{lit-arith}$ $\synt{size-prim} \rbparen$
    $\vDash \Delta\lbparen \synt{lit-arith} \rbparen \ast \Delta\lbparen \synt{size-prim} \rbparen$ \\
  $\Delta \lbparen \synt{size-arith}_l \texttt{`+'} \synt{size-arith}_r \rbparen$
    $\vDash \Delta \lbparen \synt{size-arith}_l \rbparen + \Delta \lbparen \synt{size-arith}_r \rbparen $ \\
  $\Delta \lbparen \synt{size-arith}_l \texttt{`-'} \synt{size-arith}_r \rbparen$
    $\vDash \Delta \lbparen \synt{size-arith}_l \rbparen - \Delta \lbparen \synt{size-arith}_r \rbparen $ \\

  $\Delta \lbparen \synt{field-id} \texttt{`:'} \synt{size-arith} \rbparen$
    $\vDash \Delta \lbparen \synt{size-arith} \rbparen$

\end{tabular}
  }
  \caption{\label{fig:size-rules} The rules for computing the in-memory size of
  \ourlang~arithmetic. $\mathbb{M}$ defines core expressions,
  while $\Delta$ models computations over rational numbers.
  The constants $c_w$ and $c_p$ are architecture-specific. The \texttt{int()} function
  casts a binary term to an unsigned natural number $n$.}
\end{figure}

Figure~\ref{fig:compilation-rules} shows the rules for compiling a \ourlang~surface syntax
declaration into a core \ourlang~expression. Figure~\ref{fig:size-rules} contains the definitions
for translating an arithmetic expression into natural numbers.

\addtocounter{footnote}{-2}
\stepcounter{footnote}\footnotetext{$\lbparen \dots \rbparen$ separates raised syntax (inside brackets)
from lowered expressions.}
\stepcounter{footnote}\footnotetext{A double-turnstile, Foo$\lbparen \dots \rbparen$ $\vDash$ Bar,
reads as ``Bar models Foo$\lbparen \dots \rbparen$''.}

For the translation
$\mathbb{C}\lbparen \synt{layer} \rbparen$ to be defined, a \synt{layer} must satisfy a few
properties. First, all \synt{macro} constructs must have been eliminated and syntactically
replaced with their top-level declarations. Second, the surface declaration must be validly
scoped, meaning every use of a \synt{formal-id} must be scoped inside a \synt{layer} defining
it. \ourlang~is lexically scoped with shadowing.
In Rule (1) of Figure~\ref{fig:compilation-rules} there are three optional
constructs. Each construct compiles to an expression wrapping
the compilation of the containing value: $\mathbb{C}\lbparen \synt{demarc-val} \rbparen$.
For brevity we do not show all $9$ permutations of the \synt{layer-simple} rule,
i.e. Rule (1), which represents cases where:

\begin{itemize}
  \item If \synt{formals} is missing, ``$\Exists{f_0} \cdots \Exists{f_n}$'' disappears.
  \item If \synt{mag} is missing, ``$\mathbb{M}\lbparen \synt{mag} \rbparen$'' disappears.
  \item If \synt{align} is missing, ``$@ {(\Delta_{\textrm{byte}} \lbparen \synt{align} \rbparen)}$''
    disappears.
  \item A \synt{magAlign} becomes a \synt{mag} and an \synt{align}.
\end{itemize}

%
%

\section{Rust libraries generated and results} \label{sec:rust-libraries-generated-and-resluts}

\begin{figure}
\lstset{
  basicstyle=\ttfamily\footnotesize,
  numbers=left }
\begin{lstlisting}[language=Rust,xleftmargin=5.0ex]
pub const CELL_0_OFFSET : usize = 0;
pub const CELL_1_OFFSET : usize = 8;
pub const CELL_2_OFFSET : usize = 16;
pub const CELL_3_OFFSET : usize = 24;
pub const PAYLOAD_OFFSET : usize = 32;
#[repr(C)]
#[derive(Copy, Clone, Eq, Hash)]
pub struct Cell_1Addr(usize);
pub const CELL_1_BYTES_ALIGN : usize = 1;
deriveAddr!(Cell_1Addr, CELL_1_BYTES_ALIGN);
impl Cell_1Addr {
  pub fn get_cell(self) -> CellAddr {
    self.load::<CellAddr>() }
  pub fn set_cell(self, ptr: CellAddr) {
    self.store(ptr); } }
#[repr(C)]
#[derive(Copy, Clone, Eq, Hash)]
pub struct CellAddr(usize);
pub const CELL_ALIGN : usize = 3;
deriveAddr!(CellAddr, 1 << CELL_ALIGN);
impl CellAddr {
  pub fn cell_1(self) -> Cell_1Addr {
    self.plus::<Cell_1Addr>(CELL_1_OFFSET) }
  pub fn from_cell_1(x: Cell_1Addr) -> Self {
    x.sub::<Self>(CELL_1_OFFSET) } }
\end{lstlisting}
\caption{\label{fig:rust-generated} Snippets taken from the Rust library generated
  for the immix memory layout of Figure~\ref{fig:immix-layout}.}
\end{figure}

\begin{figure}
\begin{tabular}{lllll}
$L_o$ & $L_f$ & $U_o$ & $U_f$ & File \\ \hline
24 & - & 3 & - & common/address_map.rs \\
48 & - & 3 & - & common/address_bitmap.rs \\
97 & - & 15 & - & common/bitmap.rs \\
132 & - & 9 & - & common/mod.rs \\
16 & 17 & 0 & 0 & heap/mod.rs \\
27 & 21 & 3 & 0 & objectmodel/mod.rs \\
28 & 12 & 0 & 0 & heap/immix/mod.rs \\
42 & 42 & 0 & 0 & obj_init.rs \\
51 & 53 & 2 & 1 & mark.rs \\
52 & 53 & 2 & 0 & trace.rs \\
72 & 68 & 3 & 0 & lib.rs \\
94 & 94 & 1 & 1 & heap/freelist/mod.rs \\
173 & 171 & 4 & 0 & heap/immix/immix_mutator.rs \\
222 & 224 & 8 & 2 & heap/immix/immix_space.rs \\
285 & 304 & 10 & 4 & heap/gc/mod.rs \\
- & 47 & - & 0 & heap/flp/layout.flp \\
\hline
1363 & 1107 & 63 & 8 & Total: (19\% L, 87\% U) \\ \hline \hline
- & 530 & - & 32 & heap/flp/mod.rs \\
- & 188 & - & 7 & heap/flp/address.rs \\
\end{tabular}
\caption{\label{fig:loc} Lines of immix source code, comparing the original code
of~\cite{rust-gc} with our \ourlang-integrated version. $L_o$ and $L_f$ are the
total number of non-empty lines in the original and \ourlang~version
respectively. The $U_o$ and $U_f$ columns indicate unsafe lines of code. The
subsequent two columns shows the reduction in the number of unsafe statements
in the code. Entries with a `$-$' indicate the file is not present in that
version. The \ourlang~compiler generates ``heap/\ourlangExt/mod.rs'' from the
file ``heap/\ourlangExt/layout.\ourlangExt''. We calculate line counts
ignoring blank lines, comments, and sole curly braces.}
\end{figure}

The \ourlang~language is implemented as a compiler targeting Rust code.
This section discusses the mechanics of the \ourlang~library interface,
i.e. how a \ourlang~specification integrates with a memory manager. Curious
readers should look at the \ourlang~compiler source
\fnurl{repository}{https://github.com/RedlineResearch/floorplan} to see all
the library interfaces generated. Throughout this section numbers$_{\bf 1}$
on$_{\bf 2}$ words$_{\bf 3}$ refer to line numbers in
Figure~\ref{fig:rust-generated}.



\subsection{Code generation \& library interface} \label{sec:code-generation}

Figure~\ref{fig:rust-generated} shows a sampling of the Rust library interface
generated for the immix memory layout of Figure~\ref{fig:immix-layout}.
The compiler generates a struct type for each \synt{layer-id} and \synt{field-id}.
Address types$_{\bf 8,18}$ are wrappers around a word (\verb|usize|)
with no runtime overhead. Each address type implements a Rust trait
called Address, providing a number of generic pointer and
arithmetic operations such as load$_{\bf 13}$, store$_{\bf 15}$, plus$_{\bf
23}$, and sub$_{\bf 25}$, among others\footnote{Generic access operations are
not programmer-accessible, by default.}. This trait requires
the four deriving$_{7,17}$ clauses on each address type.

Offset constants$_{1-5}$ are generated with a particular architecture in mind
(i.e. 64-bit herein).
Offset constants, along with alignment constants$_{9,19}$, are in various
places$_{10,20,23,25}$ throughout generated Rust code. The \ourlang~compiler
generates code which mimics the modularity of existing memory management
systems~\cite{dlmalloc,openjdk-hotspot,GHC-GC} and
frameworks~\cite{mmtk,HeapLayers}. This form enables pain-free manual
inspection of generated code.

Finally we have the four functions$_{\bf 12,24,22,24}$ generated in our example of
Figure~\ref{fig:rust-generated}.
The first function,
\texttt{get_cell}$_{\bf 12}$, requires a valid \texttt{Cell_1Addr} in order to
call it and returns the contents of the \texttt{cell_1} field of a cell
wrapped in a \texttt{CellAddr}.
The \ourlang~compiler and interface provide behind-the-scenes unwrapping,
accessing, and rewrapping of values (with no dynamic runtime overhead) into
Rust types. In this paradigm the Rust type system enforces address-level type
safety.
The abundance of generated Rust
address types also provided us with continual syntactic cues,
telling us which address types were involved in some computation.


The main cost we see in our approach to integrating a \ourlang~specification
with an existing garbage collector pertains to how a generated library
gets called. Upon modifying the immix specification dozens of
lines of GC code would become stale, requiring manual modifications to various
library call-sites. Such Rust compiler errors naturally provided us
with a task list of places in the GC code to update.

While integrating generated code into the immix code base we had to make a few
modest changes. The most extensive change involved modifying type signatures of
nearly every functions in the garbage collector to refer to the generated
address types. The next most extensive change involved finding each pointer
calculation in the code and replacing it with a generated version. This part
was less extensive because there were fewer pointer calculations than type
signatures in the code.
Nearly every change made involved a one-to-one replacement of individual lines.

\subsection{Results} \label{sec:results}

\begin{figure}
\begin{tabular}{@{}l|lllll}
  Benchmark & Original (s) & Floorplan (s) & GCs & Live (MB) \\ \hline
  \texttt{gcbench}  & $30.10 \pm 1.28$ & $28.94 \pm 1.84$ & $96$ & $134$ \\
  \texttt{initobj}  & $12.54 \pm 0.96$ & $12.87 \pm 1.21$ & $28$ & $114$ \\
  \texttt{exhaust}  & $15.91 \pm 0.63$ & $15.86 \pm 1.80$ & $86$ & $359$ \\
  \texttt{trace}    & $12.61 \pm 0.88$ & $12.52 \pm 0.58$ & $28$ & $114$ \\
\end{tabular}
  \caption{\label{fig:runtime} Runtimes and GCs triggered per benchmark.}
\end{figure}

In Figure~\ref{fig:loc}, we see that the programmer must write $19\%$ fewer
lines of code, including the \ourlang~specification. The first four lines of
the figure indicate the address map and bitmap files are completely eliminated
by switching to \ourlang. These files were replaced by ``layout.\ourlangExt',
just above the Total line. Most other changes were line-for-line replacements
such as changing untyped address variables into their correspondingly typed
\ourlang~address types.

In Figure~\ref{fig:loc}, we account for the number of unsafe statements of code
in the implementation before and after integrating with \ourlang. A plurality
of unsafe statements in the original code occur in special-purpose data
structures (bitmaps) which were obviated by \ourlang. In total, the number of
unsafe statements in the runtime system decreased by $87\%$: only $8$
statements remain. Of the remaining statements, four main categories emerge:
system-level allocation (2), error-handling (1), a stack-scanning FFI for C
(4), and Rust vector access optimization (1).

\ourlang~could reasonably handle system-level allocation, but we
leave this up to the programmer for increased flexibility. The
stack-scanning and error-handling lines are unsafe as a result of program
control-flow, making \ourlang~wholly unsuited to the task. Lastly, the unsafety
of an optimized vector access would seem to be suitable for \ourlang~to handle
but required converting the representation of Rust data structures
into \ourlang-constructed ones.

\noindent
{\bf Benchmarks:} We ran four benchmarks provided with the immix
implementation, respectively named exhaust, initobj, gcbench, and trace. All
benchmarks had internal parameters modified in order to trigger substantially
more GCs than originally written for, and we recorded average runtimes and
standard deviations for $100$ runs of each benchmark as detailed below.
A set of $5$ warm-up runs of each benchmark were run prior to the $100$
runs, with a $10$ second cool-down in-between benchmarks.
Benchmarks ran on a $12-$core, 2.80 GHz Intel Xeon (X5660) processor
running Arch Linux with 12 GB of RAM installed and an immix heap of $400$
MB.

The benchmarks are called gcbench, initobj, trace, and exhaust; they
respectively (1) construct application-level trees of certain depths, (2)
stress test initialization, (3) trace freshly allocated objects, and (4) induce
high memory pressure. In all cases Figure~\ref{fig:runtime} shows no
discernible difference between Floorplan's performance and the original
benchmarks, with runtimes ranging from $10-30$ seconds per run. This result
agrees with our initial hypothesis: \ourlang~generated code abstracts away
common memory layout patterns without changing the performance of address
computations.

\newcommand\myColorBox[2]{{\setlength{\fboxsep}{1.3pt}\colorbox{#1}{#2}}}
We also manually inspected the assembly code generated for accessing of
bitmaps for immix line liveness, reference bytes, and mark bits.
Figure~\ref{fig:assembly-code} shows the segment of code for line marking,
extracted from the GC's object tracing procedure. Importantly, lines $8$, $10$,
$11$, and $15$ of the original code (highlighted in \myColorBox{red!30}{red})
correspond directly to four lines in the \ourlang-generated version. Those four
lines respectively compute a byte offset of a cell into the heap$_{\bf 8}$,
compute the index of the corresponding line$_{\bf 10}$, mark the line as live
$_{\bf 11}$ ($1$ is \texttt{Live} from Figure~\ref{fig:immix-layout}), and mark
the \emph{next}$_{\bf 12}$ line as conservatively live ($3$ is
\texttt{ConservLive}).

Additionally this code detects cells outside the heap$_{\bf
1-4}$, and detects$_{\bf 12-14}$ the last line index in the
heap.\footnote{Allocating an extra entry in the line mark table would obviate
these lines.} Control-flow instructions$_{\bf 2,4,14,16}$ are highlighted in
\myColorBox{gray!20}{gray}, and the remaining instructions (in
\myColorBox{blue!40}{blue}) load metadata$_{3,5,6}$ about the heap from a Rust
struct. Modulo register allocation and precise instruction ordering, the purpose
of each line of assembly is computed with an identical instruction
opcode.

\begin{figure}
\makebox[\columnwidth]{
\newcommand\mkline[3]{%
  \tikz[baseline,every node/.style={anchor=base west}]{\node[fill=#2,anchor=base] (#1) {%
    \lstinline[basicstyle=\scriptsize\ttfamily\color{black}]|#3|};%
  }%
  \vspace{-0.5mm}%
  \\
}

\newcommand\lineno[1]{{\scriptsize #1}}

\newlength{\fulllength}
\setlength{\fulllength}{\columnwidth}
\noindent%
\begin{minipage}[t]{0.57\columnwidth}
  Original code \vspace{-3mm}\\
  \vspace{-1mm}
  \rule{\fulllength}{0.5pt}
  \lineno{\hphantom{0}1} \mkline{orig1}{blue!40}{cmp r13, rdx}
  \lineno{\hphantom{0}2} \mkline{orig2}{gray!20}{jb  .LBB250_6}
  \lineno{\hphantom{0}3} \mkline{orig3}{blue!40}{cmp r13, qword ptr [rsi + 24]}
  \lineno{\hphantom{0}4} \mkline{orig4}{gray!20}{jae .LBB250_6}
  \lineno{\hphantom{0}5} \mkline{orig5}{blue!40}{mov r8, qword ptr [rsi + 56]}
  \lineno{\hphantom{0}6} \mkline{orig6}{blue!40}{mov rbx, qword ptr [rsi + 64]}
  \lineno{\hphantom{0}7} \mkline{orig7}{blue!40}{mov rax, r13}
  \lineno{\hphantom{0}8} \mkline{orig8}{red!30}{sub rax, qword ptr [rsi + 48]}
  \lineno{\hphantom{0}9} \mkline{orig9}{blue!40}{mov byte ptr [rcx + rbp], dil}
  \lineno{10} \mkline{orig10}{red!30}{shr rax, 8}
  \lineno{11} \mkline{orig11}{red!30}{mov byte ptr [r8 + rax], 1}
  \lineno{12} \mkline{orig12}{blue!40}{add rbx, -1}
  \lineno{13} \mkline{orig13}{blue!40}{cmp rax, rbx}
  \lineno{14} \mkline{orig14}{gray!20}{jae .LBB250_6}
  \lineno{15} \mkline{orig15}{red!30}{mov byte ptr [r8 + rax + 1], 3}
  \lineno{16} \mkline{orig16}{gray!20}{.LBB250_6:}
\end{minipage}
\noindent%
\begin{minipage}[t]{0.5\columnwidth}
  Floorplan code\vspace{-3mm} \vspace{3.1mm} \\
  \mkline{flp1}{blue!40}{cmp r14, rbp}
  \mkline{flp2}{gray!20}{jb  .LBB141_6}
  \mkline{flp3}{blue!40}{cmp r14, qword ptr [rax + 24]}
  \mkline{flp4}{gray!20}{jae .LBB141_6}
  \mkline{flp5}{blue!40}{mov byte ptr [r10 + rbx], r9b}
  \mkline{flp6}{blue!40}{mov rcx, qword ptr [rax + 56]}
  \mkline{flp7}{blue!40}{mov rsi, r14}
  \mkline{flp8}{red!30}{sub rsi, qword ptr [rax + 48]}
  \mkline{flp9}{red!30}{shr rsi, 8}
  \mkline{flp10}{red!30}{mov byte ptr [rsi + rcx], 1}
  \mkline{flp11}{blue!40}{mov rax, qword ptr [rax + 64]}
  \mkline{flp12}{blue!40}{add rax, -1}
  \mkline{flp13}{blue!40}{cmp rsi, rax}
  \mkline{flp14}{gray!20}{jae .LBB141_6}
  \mkline{flp15}{red!30}{mov byte ptr [rcx + rsi + 1], 3}
  \mkline{flp16}{gray!20}{.LBB141_6:}
\end{minipage}%
\begin{tikzpicture}[overlay]
  \path[->] (orig1) edge [] (flp1);
  \path[->] (orig2) edge [] (flp2);
  \path[->] (orig3) edge [] (flp3);
  \path[->] (orig4) edge [] (flp4);
  \path[->] (orig5) edge [] (flp6);
  \path[->] (orig6) edge [out=2, in=180, out looseness=1.5] (flp11);
  \path[->] (orig7) edge [] (flp7);
  \path[->] (orig8) edge [] (flp8);
  \path[->] (orig9) edge [out=-3, in=180, out looseness=0.7] (flp5);
  \path[->] (orig10) edge [out=0, in=180, in looseness=0.95, out looseness=3.1] (flp9);
  \path[->] (orig11) edge [out=0, in=180, out looseness=1.5] (flp10);
  \path[->] (orig12) edge (flp12);
  \path[->] (orig13) edge (flp13);
  \path[->] (orig14) edge (flp14);
  \path[->] (orig15) edge (flp15);
  \path[->] (orig16) edge (flp16);
\end{tikzpicture}
}
\caption{\label{fig:assembly-code} x86 Intel assembly code for marking immix lines.}
\end{figure}

\subsection{Discussion} \label{sec:discussion}


We observe a reduction
in code-base size by nearly $20\%$ in immix-rust. This alleviates
some of the technical debt of maintaining a memory manager: eliminating numerous
interrelated offset constants and pointer arithmetic operations. These
operations corrupt memory when applied improperly. These errors could
eventually be obviated with theorem-proving techniques over Floorplan
specifications.

In lieu of obviating errors, we intend to develop debugging infrastructure capable
of detecting memory corruption at the first sign of layout integrity failure.
A layout integrity failure occurs when a load or store operation
conflicts with the addressee's intended type. The intended type of a
piece of memory derives from policy decisions made earlier in a memory
manager's execution. For example, after the mark phase of a mark-sweep garbage
collector, certain memory cells implicitly have type ``free cell''.
A buggy deallocation scheme can only corrupt memory in generated
(unsafe) address calculations. These calculations can, and we've discovered
do, encompass most all unsafe lines of code. Generated code can
readily be instrumented by the Floorplan compiler.


\section{Related work} \label{sec:related-work}


\subsection{Declarative layout specifications} \label{sec:layout-specs}

Our work is inspired by PADS~\cite{pads,forest}, a declarative embedded DSL
for describing and parsing ad hoc data structures (PADS). PADS excels
at describing log files containing textual data. For example, a PADS description
encodes arrays of partitioned data. PADS captures the structure of such an
array as a \emph{type}. \ourlang~too declaratively describes arrays of data.
In contrast to PADS, \ourlang~excels at describing heap layouts containing binary
data. A \ourlang~specification alone is not sufficient in order to
parse raw pages.


The authors of FlashRelate~\cite{flashrelate} presented work on ``a novel domain
specific language called Flare that extends traditional regular expressions
with [two-dimensional] spatial constraints.'' The underlying spatial
principle of the Flare language inspired that of \ourlang: a novel domain
specific language augmenting a context-free grammar with one-dimensional layout
constraints. The work on FlashRelate is motivated by data-cleaning tasks
and thus aims to heuristically solve the parsing of semi-structured
two-dimensional data. In contrast, this work is motivated by the runtime system
development task of implementing a memory manager and thus aims to deductively
specify the memory layout of an unstructured one-dimensional virtual address space.

\subsection{Memory management frameworks} \label{sec:frameworks}

An imperative heap layout abstraction framework known as Heap
Layers~\cite{HeapLayers} tackles the problem of implementing ``clean, easy-to-use
allocator interfaces'' which are ``based on C++ templates and inheritance.''
Heap Layers' use of template parameters is very similar to this work's notion
of declaratively specifying the properties of a memory layout. Similarly, the
Memory Management Toolkit (MMTk)~\cite{mmtk} tackles the problem of implementing
garbage collectors where the ``resulting system is more robust, easier to
maintain, and has fewer defects than monolithic collectors.''
As for defects related to memory layout, work on implementing an immix GC
in Rust~\cite{rust-gc} aims to eliminate safety defects with static safety.



\section{Conclusion} \label{sec:conclusion}

In this paper we presented a declarative language, \ourlang, for implementing the
memory layout of memory managed systems in Rust.
We presented a \ourlangLines line \ourlang~specification for the memory
layout of the state-of-the-art garbage collection algorithm immix. The
compiler generated \generatedLines lines of Rust code replacing
\pointerArithmeticLines lines of pointer arithmetic, \offsetConstantLines lines
of offset constants, and \bitmapLines lines of bitmap code.

\begin{acks}                            
This material is based upon work supported by the
\grantsponsor{GS100000001}{National Science Foundation}{http://dx.doi.org/10.13039/100000001} under Grant
No.~\grantnum{GS100000001}{1717373}.


\end{acks}


\bibliography{research.bib}

%
%
%
\end{document}